\documentclass[10pt,a4paper]{mjcs}

\usepackage{times}
\usepackage{graphicx}
\graphicspath{{./graphics/}}
\usepackage[figurename=Fig.]{caption}
\usepackage{subfig}
\usepackage[fleqn]{amsmath}
\usepackage{textcomp}
\usepackage{amssymb,amsfonts}
\usepackage{multirow}
\usepackage{color}
\usepackage{array}
\usepackage{hhline}
\usepackage{arydshln}
\usepackage{algorithm,algpseudocode}
\usepackage{etoolbox}
\usepackage{makecell}
\usepackage{stackengine}
\usepackage{pgfplots}
\usepackage[export]{adjustbox}
\usepackage{pbox}
\usepackage{pgfplots}
\usepackage{xcolor}

\pgfplotsset{width=7cm,compat=1.8}
\usetikzlibrary{patterns}
\definecolor{lightBlue}{RGB}{158,202,225}
\definecolor{darkBlue}{RGB}{50,130,189}
\definecolor{darkGreen}{RGB}{31,161,135}
\definecolor{xGreen}{RGB}{60, 180, 75}
\definecolor{xYellow}{RGB}{255, 225, 25}
\definecolor{xOrange}{RGB}{245, 130, 48}
\definecolor{xBlue1}{RGB}{0, 20, 200}
\definecolor{xBlue2}{RGB}{0, 60, 200}
\definecolor{xBlue3}{RGB}{0, 100,200}
\definecolor{xBlue4}{RGB}{0, 130, 200}
\definecolor{xBlack}{RGB}{0, 0, 09}
\definecolor{xRed}{RGB}{230, 25, 75}
\definecolor{xPurple1}{RGB}{145, 0, 130}
\definecolor{xPurple2}{RGB}{145, 30, 180}
\definecolor{xPink}{RGB}{250, 190, 190}
\definecolor{xTeal}{RGB}{0, 128, 128}
\definecolor{xNevy}{RGB}{0, 0, 120}
\definecolor{xMaroon}{RGB}{128, 0, 0}
\definecolor{xMint}{RGB}{170, 255, 195}
\definecolor{xOlive}{RGB}{128, 128, 0}
\definecolor{xCoral}{RGB}{255, 215, 180}
\definecolor{xMagenta}{RGB}{240, 50, 230}
\definecolor{xCyan}{RGB}{70, 240, 240}
\definecolor{xGray}{RGB}{128, 128, 128}

\renewcommand*\thesection{\arabic{section}.0}
\renewcommand*\thesubsection{\arabic{section}.\arabic{subsection}}
\renewcommand*\thesubsubsection{\arabic{section}.\arabic{subsection}.\arabic{subsubsection}}
\renewcommand{\refname}{REFERENCES}
\setstackEOL{\\}

\algnewcommand{\Init}[1]{%
	\State \textbf{Initialize:}
	\Statex \hspace*{\algorithmicindent}\parbox[t]{.8\linewidth}{\raggedright #1}
}
\newcolumntype{C}[1]{>{\centering\let\newline\\\arraybackslash\hspace{0pt}}m{#1}}
\def\BibTeX{{\rm B\kern-.05em{\sc i\kern-.025em b}\kern-.08em
		T\kern-.1667em\lower.7ex\hbox{E}\kern-.125emX}}
\graphicspath{ {graphics/} }

\newenvironment{mjcsabstract}{%
\begin{flushleft}
\textbf{\textit{ABSTRACT}}
\end{flushleft}

\it}

\title{\large\textbf{SELECTION OF A MINIMAL NUMBER OF SIGNIFICANT PORCINE SNPs BY AN INFORMATION GAIN AND GENETIC ALGORITHM HYBRID MODEL}}

\author
{
\textit{\textbf{Wanthanee Rathasamuth$^{1}$, Kitsuchart Pasupa$^{2,}$\footnote{Corresponding Author}{ }, Sissades Tongsima$^{3}$}}\\
\normalsize{$^{1,2}$Faculty of Information Technology,}\\
\normalsize{King Mongkut’s Institute of Technology Ladkrabang, Bangkok 10520, Thailand}\\
\normalsize{$^{3}$National Center for Genetic Engineering and Biotechnology (BIOTEC),}\\
\normalsize{National Science and Technology Development Agency (NSTDA), Pathum Thani 12120, Thailand}\\
\normalsize{E-mail: $^{1}$rathasamuth.wan@gmail.com, $^{2}$kitsuchart@it.kmitl.ac.th, $^{3}$sissades@biotec.or.th}
}

\date{}

\usepackage{fancyhdr}
 
\begin{document} 


\baselineskip12pt


\maketitle

\begin{mjcsabstract}
A panel of large number of common Single Nucleotide Polymorphisms (SNPs) distributed across an entire porcine genome has been widely used to represent genetic variability of pig. With the advent of SNP-array technology, a genome-wide genetic profile of a specimen can be easily observed. Among the large number of such variations, there exist a much smaller subset of the SNP panel that could equally be used to correctly identify the corresponding breed. This work presents a SNP selection heuristic that can still be used effectively in the breed classification process. The proposed feature selection was done by the approach of combining a filter method and a wrapper method--information gain method and genetic algorithm--plus a feature frequency selection step, while classification was done by support vector machine. The approach was able to reduce the number of significant SNPs to 0.86~\% of the total number of SNPs in a swine dataset and provided a high classification accuracy of 94.80~\%.
\end{mjcsabstract}
%
%
%
%
\section{INTRODUCTION}
\label{sec:introduction}
Swine breed improvement has played an important role in boosting the quality and quantity of pork meat in the market. Examples of swine breed that are currently popular in many countries are Landrace, Largewhite, Doroc, Creole, Wild boar, and Hampshire~\cite{burgos-paz_porcine_2013}. Each breed has distinctive characteristics. For example, the breeds that are commonly used as breeders are Largewhite, Landrace, and Duroc because they are strong, grow quickly, and provide good quality carcass, especially the Largewhite breed. The Duroc breed, on the other hand, grows well under any weather conditions and is very popular as a breeder for beautiful hybrids, while the Landrace breed is very good at rearing its offspring but carries poor traits, e.g., having weak legs. Therefore, cross breeding among these breeding stocks becomes common practice to produce desired characteristics.

The unique characteristics of each swine are manifestation of the differences in the deoxyribonucleic acid (DNA) base sequence of each swine.  DNA is a nucleic acid that stores genetic information of living beings. Unfolded, DNA can be seen as an arrangement of several nucleotides sequentially connected into two intertwining strands of polynucleotides that consist of four kinds of bases: adenine (A), Thymine (T), cytosine (C), and guanine (G). Base pairing between the two polynucleotide strands is a complementary base paring by hydrogen bonds: Adenine pairs with thymine and cytosine pairs with guanine. Two molecules of nucleotide can be arranged in DNA strands in 16 ways (or $4^n$, where \textit{n} is the number of molecules). Therefore, in a typical DNA molecule that has hundreds of thousands or even a million nucleotide pairs, the base sequence of the DNA molecules from two different beings will be very much different. It is the source of genetic polymorphism such as skin colour, height, severity of contracted disease, diverse responses to drugs. These diverse characteristics stem from different sequential base sequences which is called single nucleotide polymorphism (SNP) that can occur at any of the million positions in a DNA chain. It has been estimated that SNP can be found in every sequence of 300 bases. An example of different base sequences that result in an SNP is between GCAACGTTGA and GCAGCGTTGA. This SNP is found in more than 1~\% of individuals in a population. It is just called point mutation if it affects a smaller percentage of a population. Porcine SNP analysis can determine the SNPs that affect its reproduction and growth. However, since there are over one million SNPs in the DNA of a living being, an SNP analysis by a human expert is out of the question, notwithstanding the cost and resources needed for doing it. Therefore, a better way to address this issue is to apply bioinformatics which is an integration of biology and medical and computer sciences. Various techniques for processing data in computer are adapted for uses in bioinformatics. One of the most powerful and developed computer techniques is machine learning which integrates the sciences of computer, engineering, and statistics. Broadly speaking, this technique enables computer to respond to new data by itself based on prior information. Machine learning has been used in several branches of bioinformatics such as genomics, proteomics, microarray, systems biolog, evolution, and text mining~\cite{LarranagaMachinelearningbioinformatics2006}. This reference also describes several machine learning techniques such as Support Vector Machine (SVM), Bayesian classifiers, Decision Tree, k-Nearest Neighbours, and Artificial Neural Networks. Machine learning can be divided into three categories according to the type of learning: supervised learning, unsupervised learning, and reinforcement learning. In this work, we used the supervised learning technique which involves construction of a predictive model from a training dataset and validation of the model with a testing dataset. The algorithms used in supervised learning can be either regression or classification. In this study, it is a classification task. 

In general, a learning technique for constructing a model can support a large number of features but often is not effective at classification due to over-fitting in the cases of a much larger number of features than the number of samples. Over-fitting is the result of the constructed model having too high an accuracy which when used with a test dataset gives a low prediction accuracy. One way to solve this issue is to use a small number of features. Hence, several techniques for reducing the number of features have been proposed such as those reported in~\cite{Saeysreviewfeatureselection2007a,guyon_introduction_2003}. These review papers report applications in bioinformatics that have utilised feature selection techniques such as taxonomy, microarray domain, and mass spectrometry. They also report three types of feature selection techniques used in bioinformatics: 1) filter methods such as Euclidean distance, i-test, and information gain (IG); 2) wrapper methods such as genetic algorithm (GA) and other nature-inspired algorithms; and 3) embedded methods such as Random Forest, Weight vector of SVM, and Decision Tree.

The differences in feature selection by filter method, wrapper method, and embedded method are the following. Filter method selects features by sorting feature indexes and selecting the indices at the top rank; feature selection and classification are independent of each other here. The advantages are that it is simple and fast. Wrapper method select features by evaluating the suitability of each subset of features after classification, resulting in a subset of features that can give high classification accuracy. Since evaluation has to be done after classification, a large number of features need long computation time. Embedded method is very similar to wrapper method; the difference is that in an embedded method, feature selection is performed concurrently with classification model construction hence uses less computation time than a wrapper method. Since both wrapper and embedded methods utilise classification in the process of selection of a subset of features, they provide good learning of features and give good prediction accuracy with a training dataset. However, this good accuracy causes more over-fitting issue than filter method does when they are used to make prediction of a test dataset. 

Wrapper methods have been widely used for feature selection, especially various nature-inspired algorithms. A nature-inspired algorithm has been used to perform feature selection in~\cite{ZangReviewNatureInspiredAlgorithms2010} with the objectives of increasing the efficiency and reducing the error of prediction. In another paper, particle swarm optimisation technique was modified as Improved Binary Particle Swarm Optimisation (IBPSO) and used for selecting gene expressions in combination with k-Nearest Neighbour classifier. IBPSO can avoid getting trapped at local optimums and give good classification results~\cite{ChuangImprovedbinaryPSO2008}. In another study~\cite{HuangACObasedhybridclassification2009}, a new classifier model was proposed, hybrid Ant Colony Optimisation based classifier model, that integrated Ant Colony Optimisation technique with SVM in order to improve classification accuracy by using a small number of discriminating features. There are also several studies that proposed nature-inspired techniques for bioinformatics such as~\cite{NakamuraBinaryBatAlgorithm2013b,RodriguesBCSBinaryCuckoo2013}, and~\cite{RodriguesBinaryFlowerPollination2015} that proposed using Bat algorithm, Cuckoo search algorithm, and Flower Pollination algorithm, respectively, in combination with Optimum-Path Forest classifier. GA has been applied to many fields of study such as pattern recognition~\cite{raymer_dimensionality_2000,rokach_genetic_2008}, investigation of protein function~\cite{leijoto_genetic_2014}, and SNP selection~\cite{mahdevar_tag_2010}. In~\cite{PengMolecularclassificationcancer2003,li_robust_2005,huang_hybrid_2007}, GA was used with SVM in bioinformatics. In~\cite{lei_feature_2012-1}, GA was combined with IG to achieve better classification accuracy than each of the technique alone.

Since the swine data used in this study consisted of a large number of SNPs, the number of swine samples was small, and it was expected that only a few SNPs would affect the classification, the main objective of this study was to find and select the smallest number of SNPs that facilitated effective classification. Here, we propose feature selection by a hybrid IG+GA, a fast filter method combined with a random and selective wrapper method, which includes an SNP selection step based on the frequency of appearances in randomly-seeded datasets constructed from the whole dataset. This step is the result of our reasoning that the most relevant SNPs should be the ones that appear in most of randomly-seeded datasets. The rest of the paper is arranged as follows: Section~\ref{sec:methodology} describes the methodology including feature selection and classification techniques; Section~\ref{sec:dataset} describes the dataset used in this study; Section~\ref{sec:expSetup} describes the experimental setup; Section~\ref{sec:experiment} is the results and discussion section; and section~\ref{sec:conclude} is the conclusion of the paper.
%
\section{METHODOLOGY}
\label{sec:methodology}
%
Described in this section are the conceptual framework of this study and feature selection by a hybrid IG and GA (IG+GA) technique. This technique is proposed to solve an issue that many features of the whole features are not significant to construction of a learning model but may waste computer resources and lengthen computation time. The technique was intended to select the minimum number of most significant features that can classify SNPs accurately. IG, GA, IG+GA, SVM (the classifier) are explained briefly along with feature selection according to their frequency of appearance in randomly-seeded datasets.
%
\subsection{Information gain}
\label{method:IG}
%
IG is a feature selection technique of filter method type~\cite{Saeysreviewfeatureselection2007a,guyon_introduction_2003,LazarSurveyFilterTechniques2012} that selects features according to the ranked index weights calculated from the relationships between features. It is a very popular feature reduction technique that can boost classification capability of any classifiers. It has been applied to applications such as feature selections of text, DNA microarray, and SNPs. In~\cite{ChoMachineLearningDNA2003}, it was used to select features of DNA microarray. The results from that study show that IG and Pearson’s correlation coefficient were the best among all the techniques tested including Multi-layer Perceptron and k-Nearest Neighbour. In~\cite{Jirapech-UmpaiFeatureselectionclassification2005}, six filter methods were used to rank features and three cut-point determination methods were used to find a good cut-point. It was found that IG was the best feature ranking method and Z-score analysis was the best cut-point determination method. Using these two methods in combination, the microarray was classified with the highest accuracy.

The IG value of each feature is calculated from the difference between the initial information entropy and the current information entropy of the feature. The difference is between 0 and 1. Entropy is a measure of unpredictability of the state, or equivalently, of its average information content. An information entropy signifies the difference between data points: a higher entropy means that the data points are very much different while a lower entropy means that the data points were not very different. Therefore, a feature with a high IG value is a good feature. Calculation of IG value is expressed in Eq.~(\ref{eqn:1}) below,
\begin{equation} 
\label{eqn:1}
IG(T,i)=H(T)-\sum v\in vals(i)\frac{|\{\textbf{\textit{x}}\in T\mid x_i=v\}|}{|T|}\cdot H(\{\textbf{\textit{x}}\in T\mid x_i=v\}),
\end{equation}
and $H$ is information entropy that can be calculated by
\begin{equation} 
\label{eqn:2}
	H(T)=-\sum_{x\in \textbf{\textit{x}}}p(x)\log_2p(x)
\end{equation}
where $T$ is training dataset with samples in the form of $(\textbf{\textit{x}},y)=\{x_1,x_2,\dots,x_k,y\}$ where $ x_i $ is the feature at the present position $i$ of the sample and  $y$ is the corresponding class label of the $\textbf{\textit{x}}$ sample; and $p(x)$ is the proportion of the number of elements in sample $\textbf{\textit{x}}$ to the number of elements in set $T$.

IG can rank features according to their significance but cannot determine the optimum number of features for a classification purpose. In this study, an elbow method was used to reliably determine the cut-point, i.e., the number of highest-ranked features sorted by IG, that would be optimum for classification purpose. The elbow method is a method of interpretation and validation of consistency within cluster analysis designed to help finding the appropriate number of clusters in a dataset.
%
\subsection{Genetic algorithm}
\label{method:GA}
%
GA is one of nature-inspired algorithms. As a feature selection technique, it falls under the wrapper method category. GA mimics the evolution process in nature and genetic inheritance in its search for a solution of an optimisation problem. It crosses over solutions then select better solutions that are represented as chromosomes that contain several genes. In GA, chromosomes are in the form of strings of alphabets or binary bits. In recent years, GA has been used for reducing the number of dimensions of data in pattern recognition process~\cite{raymer_dimensionality_2000,rokach_genetic_2008}. As mentioned above, for feature selection process, too many features but too small number of samples will degrade the performance of machine learning process. There have been several studies on feature selection that propose various techniques to solve this issue~\cite{ZangReviewNatureInspiredAlgorithms2010,ChuangImprovedbinaryPSO2008,HuangACObasedhybridclassification2009}. GA is widely used for feature selection~\cite{PengMolecularclassificationcancer2003,li_robust_2005,huang_hybrid_2007}. It was combined with SVM and used in a bioinformatic application~\cite{PengMolecularclassificationcancer2003}, classification of array-based multiclass tumor. It was also used for predicting protein function in~\cite{leijoto_genetic_2014} where GA was used to select some variables before they were used further by SVM. Their prediction results were compared to those obtained by Borro et al.~\cite{BorroPredictingenzymeclass2006b} and found to be clearly better, demonstrating that using GA to select a small number of significant variables was more effective that the technique used by Borro et al. In~\cite{Ilhangeneticalgorithmsupport2013}, GA was used to find the optimum parameters, including hyper-parameters, for SVM operation.

\begin{figure}[htbp]
	\centering
	\includegraphics[width=0.5\linewidth]{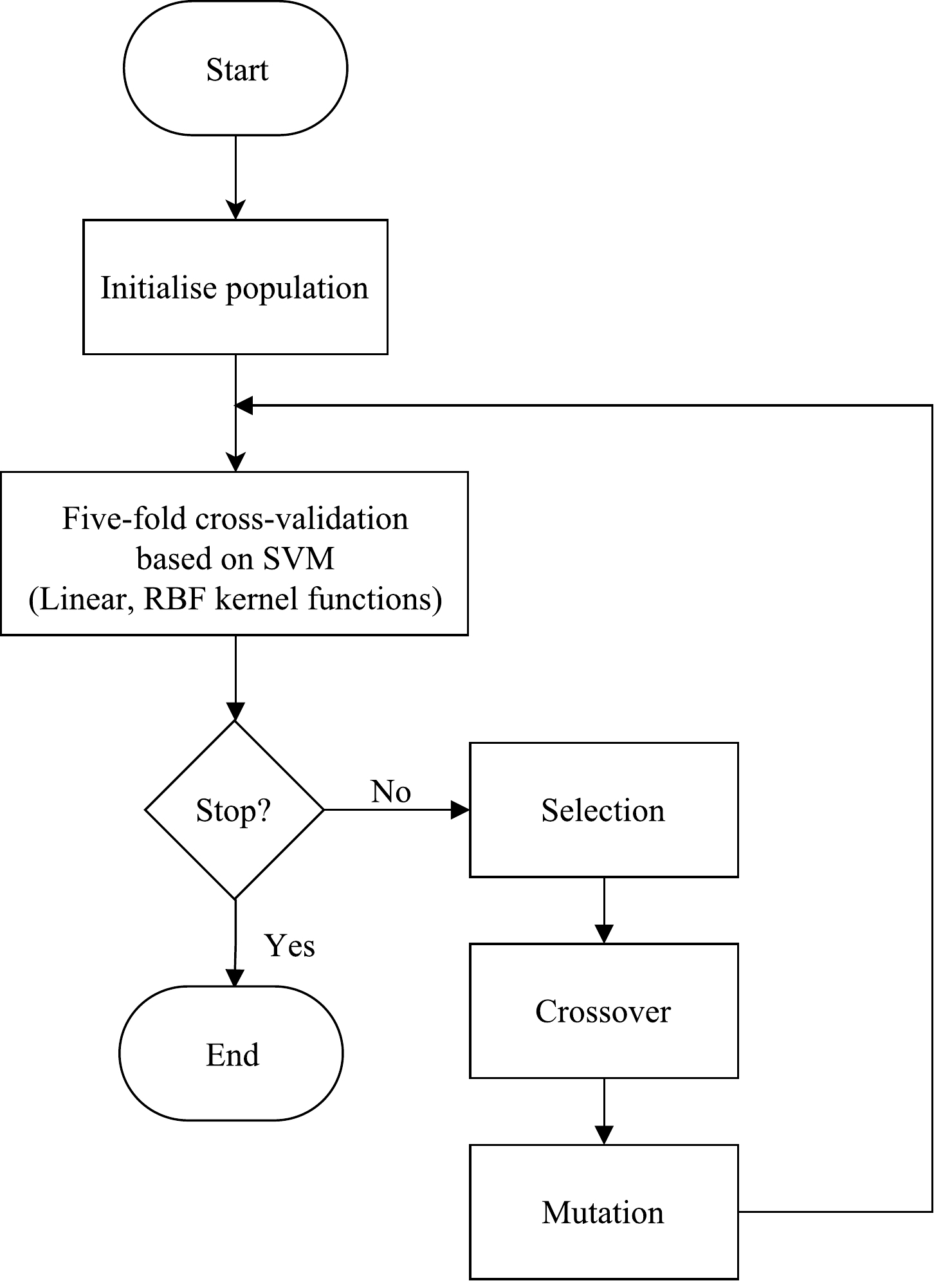}
	\caption{Procedural steps of GA in operation with SVM.}
	\label{fig:01} 
\end{figure}

\begin{figure}[htbp]
	\centering
	\includegraphics[width=0.8\linewidth]{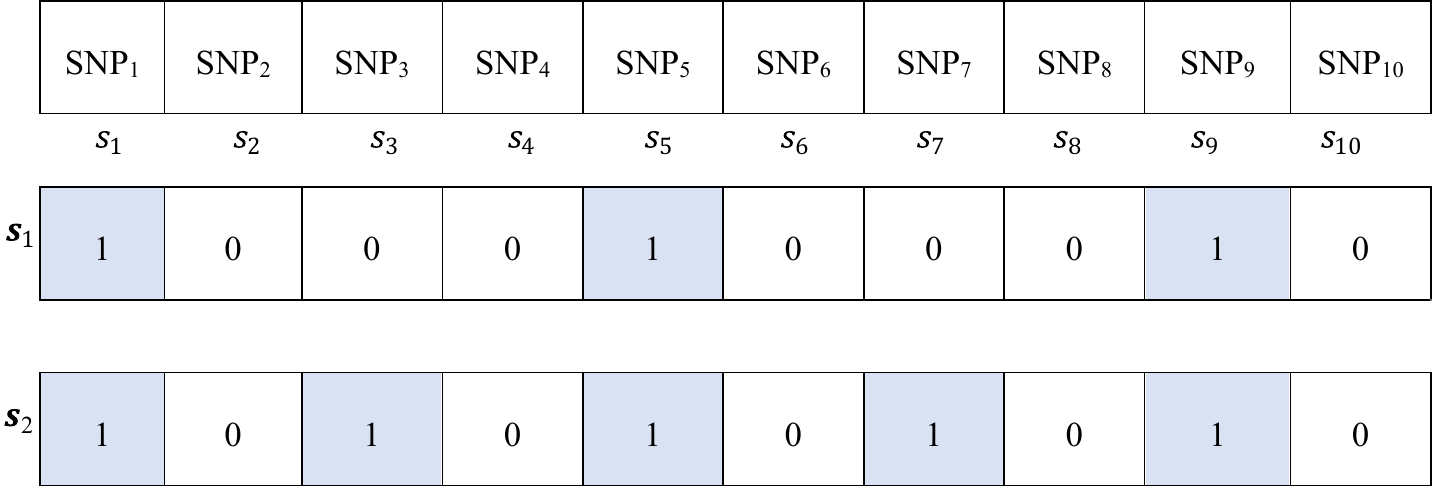}
	\caption{An example of binary bit strings of genes that make up two chromosomes.}
	\label{fig:02} 
\end{figure}

GA procedure consists of specifying the number of chromosomes with their gene components in the population, specifying their fitness function for the evolution process, populating the procedure with a random initial population, applying genetic operators--selection, crossover, and mutation--to the population, then repeating these steps to the new population until the stopping criterion is met. Fig.~\ref{fig:01} shows GA procedural steps specifically for feature selection. These steps are explained in detail below.
%
\begin{enumerate}
	\item {Generation of an initial random population of chromosomes that are binary bit strings: Each chromosome \textit{\textbf{s}} consists of \textit{n} genes $s$ where $\textbf{\textit{s}}=\{s_1 ,s_2  ,\dots,s_n\}$. Fig.~\ref{fig:02} shows an example of chromosomes represented by bit strings. $\textbf{\textit{s}}_1$ and $\textbf{\textit{s}}_2$ are two chromosomes each containing 10 binary bit strings of genes ($n = 10$) that represent 10 features in the sense that a binary value 1 at a position in the string means that the corresponding feature in the training dataset is selected for fitness function evaluation to find out which chromosome is the best one. In this study, the fitness function was accuracy. In Fig.~\ref{fig:02}, $\textbf{\textit{s}}_1$ and $\textbf{\textit{s}}_2$ have three and five selected SNPs, respectively.}
	\item {Selection of to-be-reproduced chromosomes by roulette wheel method: The roulette wheel selection method selects a chromosome randomly based on its selection probability which is the ratio of its fitness value to the total fitness value of the entire population.}
	\item {Crossover of two chromosomes: it is done by exchanging some of their genes to get new chromosomes that may be better than the original ones. The crossover procedure is a multi-point crossover that starts with generation of random numbers that specify the positions and blocks of genes that will be crossed over.}
	\item {Mutation of the cross-overed chromosomes: mutation is done to increase the diversity of chromosome population. Even though the selection and crossover operators may give better solutions, the solutions are still based on the original chromosomes and so may not be diverse enough to reach a global optimum. Mutation is a procedure that can generate diverse solutions that may not be obtainable from the original information stored in the parent chromosomes. The first mutation method used in this study was bit-flip mutation that is the method used in the original formulation of GA. Bit flipping is based on mutation probability $P_m$. For instance $P_m$ equals to 0.01 means that the bit representing the gene has a 1~\% chance to flip from 0 to 1 or 1 to 0.}
\end{enumerate}
%
However, it was found that after the mutation step, the number of 1s in each mutated chromosomes was still too high, around 50~\% of all of the genes, and this could cause a problem of getting too many eligible features. In order to obtain a smaller number of optimum features, we used a higher probability value for 1 to 0 bit-flipping than that for 0 to 1 bit-flipping as shown in Eq.~(\ref{eqn:3})~\cite{mahdevar_tag_2010} which had proven to be successful. For example, $P_m$ equals to 0.1 means that the bit representing the gene has a 10~\% chance to flip from 0 to 1 and 1 to 0 was much higher at 90~\%.
\begin{equation} 
\label{eqn:3}
s(i)=\begin{cases}1 & ; r\le P_m\\0 & ;Otherwise\end{cases}
\end{equation}
Eq.~\ref{eqn:3} includes the conditions for bit-flipping of particular existing bit values: where $s(i)$ is the flipped bit at $i$, and $r$ is a random number between 0 and 1.
%
\subsection{Information gain and genetic algorithm hybrid}
\label{method:Hybrid}
%
In a previous study~\cite{lei_feature_2012-1}, IG+GA was used for text classification. Information gain calculated how many terms can be used for classification of information, in order to measure the importance of the lexical items for classification. Subsequently, GA was used to select the most suitable features. 

GA alone could not reduce the number of features sufficiently in the SNP-feature-reduction tests that we ran and even though IG+GA could reduce the number of features to a minimum, those features did not result in accurate predictions due to too small a number of features. Consequently, we chose to employ a different approach for combining IG+GA for our classification task where IG was used to rank features according to their significance; an elbow method was used to find a cut-point for inclusion of only some of the features obtained from IG which also specified the number of genes in each chromosome in subsequent Proposed GA; and GA was used to further reduce the number of these features down to a suitable number by adjusting the mutation probabilities for 0 to 1 bit-flipping and 1 to 0 bit-flipping separately. A suitable number of features here means that they provided in a good classification accuracy in test runs.

%
\subsection{Support vector machine}
\label{method:SVM}
%
SVM is a machine learning technique of the supervised learning category. It was developed to solve binary classification problems. The main concept of this technique is hyperplane construction. In SVM, a hyperplane is a decision plane for dividing data into two classes. An optimum hyperplane has the largest margin between the two classes. The data on the margin are called support vectors. SVM can have one of many kinds of kernel functions such as linear, radial basis function (RBF), and polynomial kernels. These different functions map data from input space to feature space with higher dimensions. Each kernel function is appropriate for a different kind of problems: the function used needs not always be linear, depending on the type and complexity of the input data. SVM has been applied as a classifier in several research studies~\cite{PengMolecularclassificationcancer2003,li_robust_2005,huang_hybrid_2007}. In this study, linear and RBF were tested and compared of their performances. For the test, $C$ is a hyperparameter of SVM that balances training error and model's complexity. For the RBF kernel especially, a parameter $\gamma$ was tuned to get the optimal hyperplane. The optimal parameters were validated by a five-fold cross-validation procedure. The respective mathematical expressions for linear kernel function and RBF are in Eq.~(\ref{eqn:4}) and~(\ref{eqn:5}), respectively.
\begin{equation} 
\label{eqn:4}
k(x,x') = x^T\cdot x'
\end{equation}
\begin{equation} 
\label{eqn:5}
k(x,x') = \exp(-\gamma ||x-x'||^2)
\end{equation}
where $k(x,x')$ is a kernel function;  $x$ and $x'$ are data samples; the term $||(x-x')||^2$ is a squared Euclidean distance between $x$ and $x'$; and $\gamma$ is a non-negative value.

\begin{figure}[htbp]
	\centering
	\includegraphics[width=0.8\linewidth]{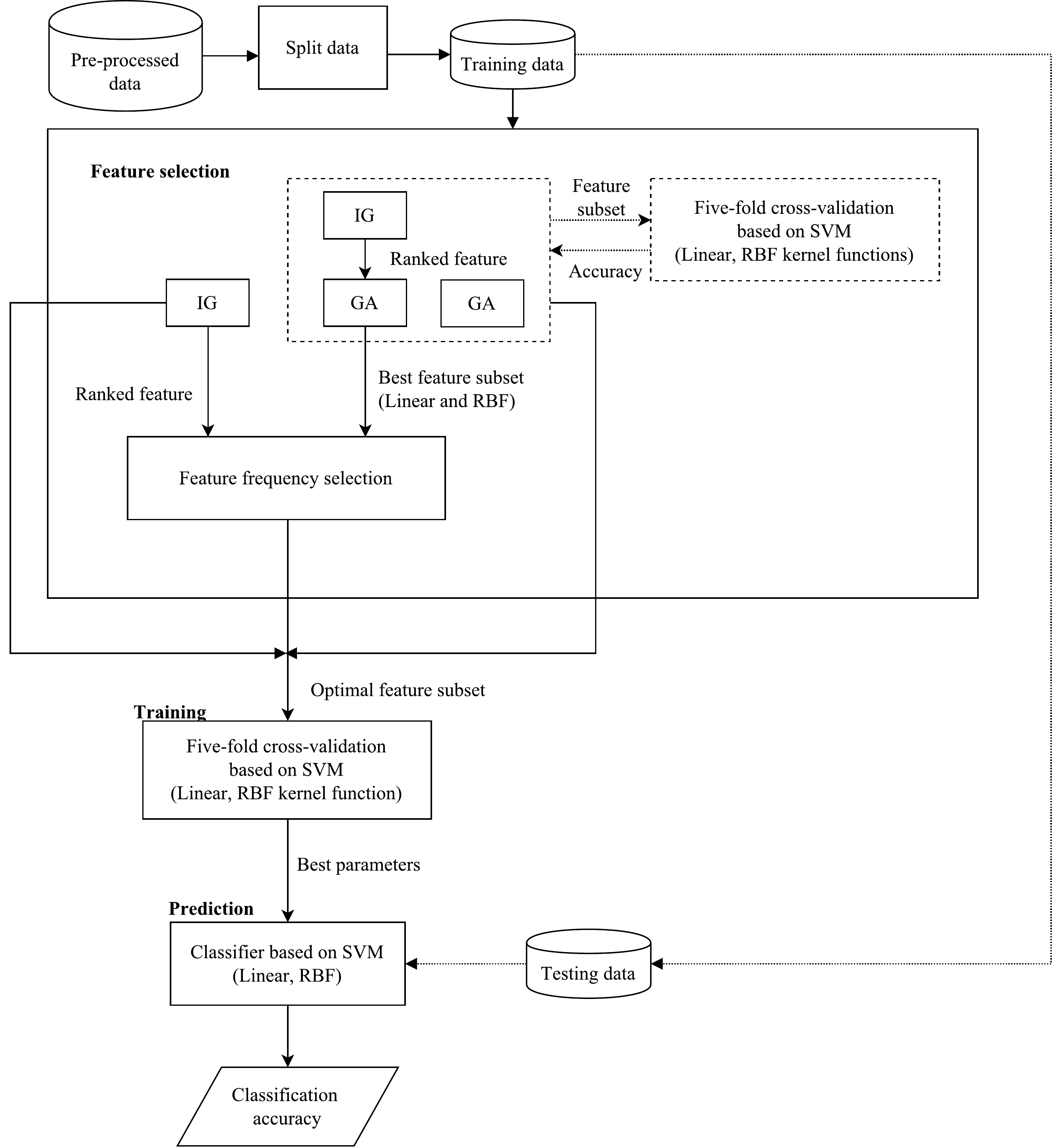}
	\caption{An experimental framework of feature selection for classification.}
	\label{fig:03} 
\end{figure}

A diagram of feature selection by our proposed approach is shown in Fig.~\ref{fig:03}. In this approach, we also compared the effectiveness of IG and GA alone as well as of IG+GA, hence their uses are as shown in the diagram. As mentioned in the section above, we introduced different flipping probabilities of 1 and 0 into GA; therefore, from now on we will call this GA as a ``Proposed GA''. Please note that both GA and Proposed GA were used individually and in combination with IG. This is not shown explicitly in the diagram.

\begin{figure}[htbp]
	\centering
	\includegraphics[width=0.5\linewidth]{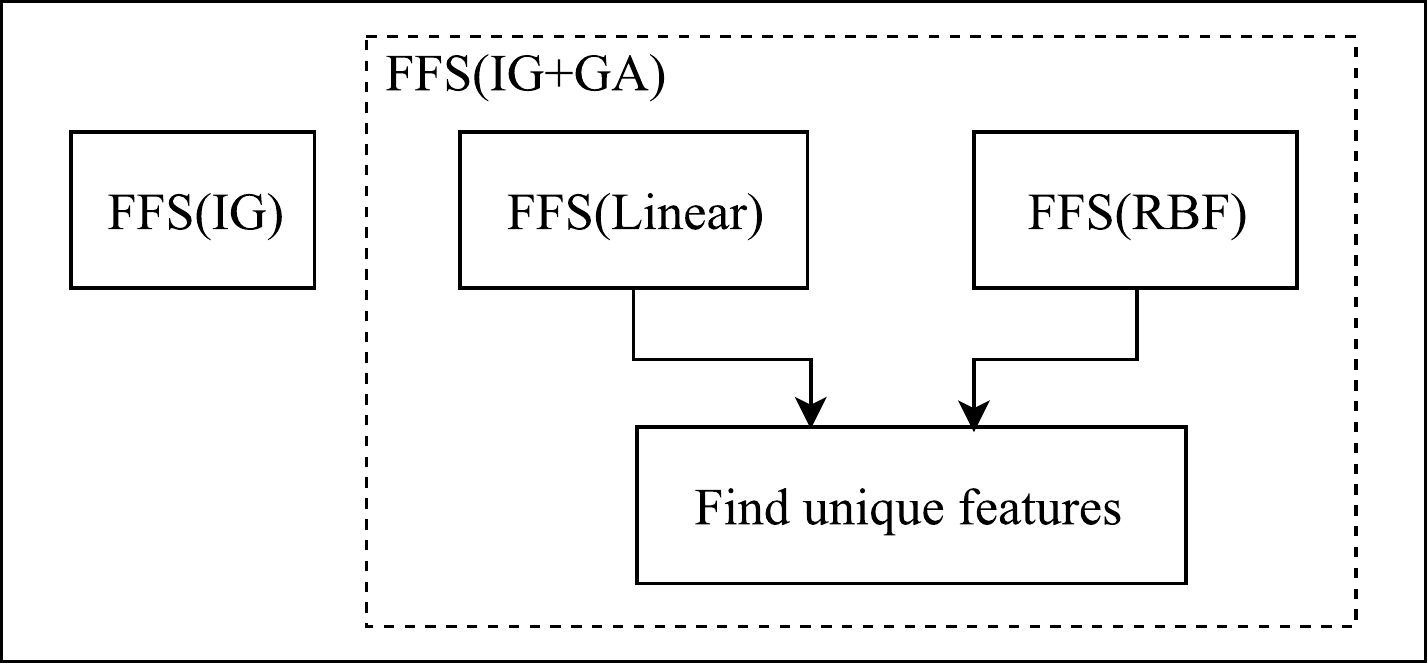}
	\caption{Application of FFS to combining and selecting features from linear and RBF kernels.}
	\label{fig:04} 
\end{figure}

As can be seen in the diagram, firstly in this experiment, pre-processed data were divided into two sets: training and test sets. The training dataset was used in feature selection. In our experimental framework, three feature selection methods were evaluated: filter, wrapper, and a combination of filter and wrapper. 
\begin{enumerate}
\item {The filter method used IG for ranking the level of significance of each feature and an elbow method to find the cut point for selection of the optimal number of features.}
\item {The wrapper method used GA for selection of an optimal subset of features for classification. To find this optimal subset, GA needed to send a preliminary subset of features into the classification process used for training, testing to find optimal parameters and evaluating the SVM model by five-fold cross validation. The best subset of features gave the highest prediction accuracy.}
\item {The filter plus wrapper combination method performed the filter and wrapper methods in that order. The cut point from the elbow method in the filter method would set the number of genes in each chromosome to be performed in GA.}
\end{enumerate}
In the development of our approach, we made an assumption that from all 10 random-seeded datasets, it was likely that some features from every dataset would be repeatedly selected. Thus, their high frequency of occurrences in the selected features meant that they were the most significant features. Therefore, we introduced a feature frequency selection (FFS) step after the selected features from IG, IG+GA, and IG+Proposed GA were obtained in order to select only a small number of the most significant features. Briefly, FFS works to find the previously selected features that have the highest frequency of occurrences among the random-seeded datasets. In this study, since IG+GA+FFS and IG+Proposed GA+FFS used both linear and RBF kernels, FFS was separately performed on the features selected by each of these kernels. the newly selected features from both kernels were then combined and the same ones were taken to be the finally-selected features as shown in Fig.~\ref{fig:04}.

After features were selected, they are used in a training step, through five-fold cross-validation, to find the optimal parameters for constructing the optimal model. Specifically for RBF kernel in SVM, grid search method is used to find the best $C$ and $\gamma$ for the SVM model. The optimal parameters are input into the prediction step together with the test dataset. The output is classification accuracy.
%
\section{DATASET}
\label{sec:dataset}
%
All swine data used in this study were from the Porcine Colonisation of the Americas Dataset~\cite{burgos-paz_porcine_2013}. It consists of data from 11 village pig breeds including Creole, Moura, Yucatan, Ossabaw pig, Monterio, and Guinea hog that are raised in the United States of America as well as 10 outgroup pig breeds including Jiangquhai, Jinhua, Meishan, Xiang pig, Duroc, Landrace, and Largewhite. The dataset contains data from 389 pig samples and 46,259 SNPs which was gleaned by a PLINK method from 62,163 SNPs. Some of the breeds presented in the dataset had too few samples representing them; therefore, those breeds were excluded from the study. In total, the dataset that we used in this study consisted of data from 356 samples of 21 breeds as shown in Table~\ref{table:1}, and a total of 16,579 SNPs. All data were put through data cleansing according to the principle of population and sample identification. However, there were some missing values; thus, they were estimated by a single imputation method. The estimated values were modes of the whole individual feature data.
\begin{table}[htbp]
	\centering
	\caption{An example of swine data in the dataset used in this study.}
	\label{table:1}
	\begin{tabular}{|l|p{9.5cm}|c|}
		\hline
		\textbf{Breed} & \textbf{Location} & \textbf{Number of samples} \\ \hline
		Creole 
		& Alto Baudo-Colombia, Baja Verapaz-Guatemala, Granma-Cuba,Guanacaste, Alajuela-Costa Rica, Loja-Ecuador,  Misiones-Argentina, Pinar del Rio-Cuba, Titicaca area-Peru 
		& 90 \\ \hline
		Piau & Bahia-Brazil & 9 \\ \hline
		Zungo & Cerete-Colombia & 10 \\ \hline
		Jiangquhai & China & 11 \\ \hline
		Jinhua & China & 16 \\ \hline
		Meishan & China & 16 \\ \hline
		Xiang pig & China & 11 \\ \hline
		Moura & Concordia-Brazil & 9 \\ \hline
		Duroc & Denmark, Holland, USA & 20 \\ \hline
		Landrace & Denmark, Holland, USA & 20 \\ \hline
		Largewhite & Denmark, Holland, USA & 20 \\ \hline
		Semi- feral & Formosa-Argentina & 10 \\ \hline
		Wild boar & Hungary, Poland, Tunisia & 13 \\ \hline
		Yucatan & Indiana-USA & 10 \\ \hline
		Hairless & Mexico & 9 \\ \hline
		Cuino & Nayarit-Maxico & 7 \\ \hline
		Ossabaw pig & Ossabaw island-USA & 7 \\ \hline
		Monteiro & Pocone-Brazil, Portugal & 24 \\ \hline
		Iberian & Spain & 15 \\ \hline
		Hampshire & UK, USA & 14 \\ \hline
		Guinea hog & USA & 15 \\ \hline
	\end{tabular}
\end{table}
%
%
\section{EXPERIMENTAL SETUP}
\label{sec:expSetup}
%
The entire swine dataset was used to construct 10 random-seeded datasets. This large number of random-seeded datasets was used in order to make the results of this experiment statistically valid and reliable. Each random-seeded dataset was partitioned into a training dataset (80~\%) and a testing dataset (20~\%). The parameter settings of GA, Proposed GA, IG+GA, and IG+Proposed GA were the following: population size of 30 chromosomes; crossover probability of 0.8; $P_m$ ranging from 0.1 to 0.9; the number of genes of 16,579 for GA and Proposed GA; the number of generations of 10; $C$ ranging from $10^{-6}$ to $10^6$; and $\gamma$ of RBF ranging from $10^{-10}$ to $10^{10}$. In FFS for IG, IG+GA, and IG+Proposed GA, features with frequency over 80~\% and higher were selected, i.e., features that occurred greater than or equal to eight times in the 10 random-seeded datasets. 

The reason that we set the population size to be a small number of 30 was that a higher number would result in a large number of features which would have wasted a lot of computational time. In addition, the reason that we set the number of generations to be 10 was that preliminary trial runs showed that GA met its stop criteria within 10 generations, so setting it to a higher number was not likely to increase the accuracy in any way. The full experimental results are reported in Section 5.0 below.
%
\section{EXPERIMENTAL RESULTS AND DISCUSSION}
\label{sec:experiment}
%
In this section the result of SNPs selection for classification are presented and discussed.  Besides presenting the average accuracy of classification by the proposed approach and the numbers of finally selected SNPs by every method used in this approach, we also present the average accuracy of classification from using the entire features for comparison. All of these results are displayed in Fig.~\ref{fig:05} and~\ref{fig:09}. This section also presents the statistical results of ANOVA analysis of the prediction accuracy achieved by every method and the results of a principal component analysis (PCA).

In our Proposed GA, mutation probability refers to the frequency of new mutations per generation in an organism or a population. It was observed that from the range of $P_m$ (0.1-0.9) set for running GA, Proposed GA, IG+GA, and IG+Proposed GA, the values of the best-tuned $P_m$ for those methods were 0.8, 0.2, 0.5, and 0.8, respectively, as shown in Fig.~\ref{fig:05}, for each value of the $P_m$ tested, GA, IG+GA, and IG+Proposed GA gave nearly the same number of selected SNPs as shown in Fig.~\ref{fig:05-a},~\ref{fig:05-c} and~\ref{fig:05-d}, so the optimal value of $P_m$ was considered to be the value that provided the highest classification accuracy. On the other hand, for Proposed GA, the number of selected SNPs obtained from using different values of $P_m$ were not nearly the same as shown in Fig.~\ref{fig:05-b} but the number of selected SNPs from $P_m$ of 0.1 and 0.2 were the lowest and nearly the same. Classification accuracy from $P_m$ of 0.2 was higher than that from $P_m$ of 0.1, so it was used as the optimal $P_m$ for Proposed GA. To conclude, it can be seen that for GA and IG+GA that used the original type of bit-flip probability, no matter what $P_m$ value was used, the total number of selected SNPs was large, about 50~\% of search space, but for Proposed GA and IG+Proposed GA that used our proposed type of bit-flip probability, the total number of selected SNPs was smaller but the optimum number depended on the size of the search space: a large search space needed a smaller value of $P_m$, but a small search space needed a larger value of $P_m$. The results presented here are the results from using these tuned values of $P_m$ with the respective methods.

\begin{figure}[]
	\centering
\begin{tikzpicture}
font=\footnotesize
\begin{axis}[
ybar=0pt,
legend image code/.code={
	\draw [#1] (0cm,-0.1cm) rectangle (0.25cm,0.1cm);
},
width=11.0cm,height=6cm, 
axis x line*=bottom,
axis y line*=left,
ytick align=outside,
bar width=0.8cm, 
enlarge x limits=0.2,
legend style={at={(0.15,1)},anchor=north,legend columns=-1},
ylabel near ticks,
ymajorgrids=true,
xtick=data,
ylabel=Number of generations,
xlabel=Methods,
xticklabels = {GA,Proposed GA,IG+GA,IG+Proposed GA},
]
\addplot[black,fill=lightBlue] table[x index=0,y index=1] \datayy; 
\addplot[black,fill=darkBlue] table[x index=0,y index=2] \datayy; 
\legend{Linear,RBF}
\end{axis}
\end{tikzpicture}
\caption{The number of generations at the stop of runs of all 10 randomly-seeded datasets of each tested method.}
\label{fig:009} 
\end{figure}
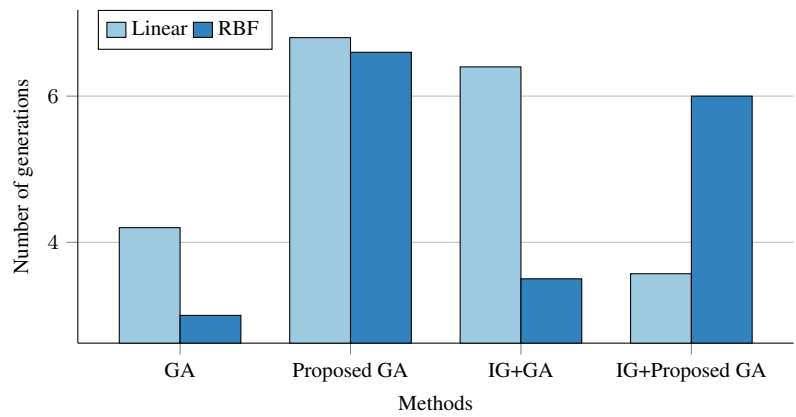

The highest levels of classification accuracy from the training step of GA, Proposed GA, IG+GA and IG+Proposed GA are shown in Fig.~\ref{fig:08}. It can be seen that the number of generations at the stop of a run of the first set of randomly-seeded dataset of each tested method was not over 10. In addition, IG+GA and IG+Proposed GA gave better levels of accuracy than GA and Proposed GA alone. The average number of generations at the stop of runs of all 10 randomly-seeded datasets of each tested method was between 3 and 7 generations, as shown in Fig.~\ref{fig:009}.
%
\pgfplotstableread{
	0	92.61538462	92.46153846
	1	92.61538462	92.61538462
	2	92.76923077	92.61538462
	3	92.76923077	92.76923077
	4	92.30769231	92.30769231
	5	92.46153846	92.61538462
	6	92.61538462	92.61538462
	7	92.92307692	92.46153846
	8	92.76923077	92.61538462
}{\dataALeft}
\pgfplotstableread{
	0	91.69230769	92
	1	91.69230769	92.15384615
	2	92.15384615	91.69230769
	3	92.76923077	92.30769231
	4	92.61538462	92.61538462
	5	92.61538462	92.92307692
	6	92.76923077	92.61538462
	7	92.92307692	92.76923077
	8	92.92307692	92.76923077
}\dataBLeft
\pgfplotstableread{
	0	91.07692308	91.23076923
	1	91.23076923	90.61538462
	2	90			91.07692308
	3	91.07692308	90.92307692
	4	91.07692308	91.84615385
	5	90.15384615	88.92307692
	6	89.53846154	89.53846154
	7	89.53846154	89.23076923
	8	89.38461538	90.92307692
}\dataCLeft
\pgfplotstableread{
	0	91.23076923	91.69230769
	1	91.23076923	91.69230769
	2	91.23076923	91.69230769
	3	90			90.76923077
	4	89.07692308	89.23076923
	5	89.69230769	89.84615385
	6	91.69230769	91.23076923
	7	92.15384615	91.84615385
	8	90.92307692	91.84615385
}\dataDLeft
%
\pgfplotstableread{
	0	8301.7	8283.9
	1	8242.6	8253.1
	2	8242.6	8249.8
	3	8275.8	8274.9
	4	8264.3	8296.1
	5	8255.9	8229.9
	6	8266.8	8264.6
	7	8239.8	8245.1
	8	8272	8280.7
}{\dataARight}
\pgfplotstableread{
	0	839.8	1496
	1	1176.5	1113.1
	2	2247.9	2158.8
	3	3944.6	3801.2
	4	5621.2	5711.4
	5	7934	7892.6
	6	9718.9	9653.6
	7	10209.4	9866.8
	8	9417.9	8246.9
}\dataBRight
\pgfplotstableread{
	0	162.8	162.4
	1	156.6	163.9
	2	160.7	155.4
	3	163		168.5
	4	165.6	165.3
	5	167.9	159.3
	6	166.8	163.1
	7	170.3	156.1
	8	169.1	159.9
}\dataCRight
\pgfplotstableread{
	0	163.3	164.8
	1	163.3	164.8
	2	163.3	164.8
	3	157.5	156.4
	4	127.3	126.7
	5	153.9	153.4
	6	197.3	197.2
	7	238.1	241.3
	8	279.3	284.9
}\dataDRight
\begin{figure}[htbp]
	\centering
	\captionsetup[subfloat]{aboveskip=-5em,belowskip=-1em}
	\subfloat[\label{fig:05-a} GA]{
		\begin{tikzpicture}
		font=\footnotesize 
		\begin{axis}[
		ybar=0pt,
		legend image code/.code={
			\draw [#1] (0cm,-0.1cm) rectangle (0.25cm,0.1cm);
		},
		ylabel near ticks,
		width=8.0cm,height=5.5cm,
		axis x line*=bottom,
		axis y line*=left,
		ytick align=outside,
		xtick=data,
		enlarge x limits=0.1,
		bar width=0.27cm, 
		legend style={at={(0.15,1)},anchor=north,legend columns=-1},
		ymajorgrids=true,
		ylabel={Accuracy (\%)},
		xlabel=Probability of mutation,
		xticklabels = {0.1,0.2,0.3,0.4,0.5,0.6,0.7,0.8,0.9},
		]
		\addplot[black,fill=lightBlue] table[x index=0,y index=1] \dataALeft; 
		\addplot[black,fill=darkBlue] table[x index=0,y index=2] \dataALeft; 
		\legend{Linear,RBF}
		\end{axis}
		\end{tikzpicture}
		\hspace{0.3cm}
		\begin{tikzpicture}
		font=\footnotesize
		\begin{axis}[
		ybar=0pt,
		width=8.0cm,height=5.5cm, 
		axis x line*=bottom,
		axis y line*=left,
		ytick align=outside,
		bar width=0.27cm, 
		enlarge x limits=0.1,
		ylabel near ticks,
		ymajorgrids=true,
		xtick=data,
		ylabel=Number of SNPs,
		xlabel=Probability of mutation,
		xticklabels = {0.1,0.2,0.3,0.4,0.5,0.6,0.7,0.8,0.9},
		]
		\addplot[black,fill=lightBlue] table[x index=0,y index=1] \dataARight; 
		\addplot[black,fill=darkBlue] table[x index=0,y index=2] \dataARight; 
		\end{axis}
		\end{tikzpicture}
	}
	\\
	\subfloat[\label{fig:05-b} Proposed GA]{
		\begin{tikzpicture}
		font=\footnotesize 
		\begin{axis}[
		ybar=0pt,
		ylabel near ticks,
		width=8.0cm,height=5.5cm,
		axis x line*=bottom,
		axis y line*=left,
		ytick align=outside,
		xtick=data,
		enlarge x limits=0.1,
		bar width=0.27cm, 
		legend style={at={(0.15,1)},anchor=north,legend columns=-1},
		ymajorgrids=true,
		ylabel={Accuracy (\%)},
		xlabel=Probability of mutation,
		xticklabels = {0.1,0.2,0.3,0.4,0.5,0.6,0.7,0.8,0.9},
		]
		\addplot[black,fill=lightBlue] table[x index=0,y index=1] \dataBLeft; 
		\addplot[black,fill=darkBlue] table[x index=0,y index=2] \dataBLeft; 
		\end{axis}
		\end{tikzpicture}
		\hspace{0.3cm}
		\begin{tikzpicture}
		font=\footnotesize
		\begin{axis}[
		ybar=0pt,
		width=8.0cm,height=5.5cm, 
		axis x line*=bottom,
		axis y line*=left,
		ytick align=outside,
		bar width=0.27cm, 
		enlarge x limits=0.1,
		ylabel near ticks,
		ymajorgrids=true,
		ymin=0,
		xtick=data,
		ylabel=Number of SNPs,
		xlabel=Probability of mutation,
		xticklabels = {0.1,0.2,0.3,0.4,0.5,0.6,0.7,0.8,0.9},
		]
		\addplot[black,fill=lightBlue] table[x index=0,y index=1] \dataBRight; 
		\addplot[black,fill=darkBlue] table[x index=0,y index=2] \dataBRight; 
		\end{axis}
		\end{tikzpicture}
	}
	\\
	\subfloat[\label{fig:05-c} IG+GA]{
		\begin{tikzpicture}
		font=\footnotesize 
		\begin{axis}[
		ybar=0pt,
		ylabel near ticks,
		width=8.0cm,height=5.5cm,
		axis x line*=bottom,
		axis y line*=left,
		ytick align=outside,
		xtick=data,
		enlarge x limits=0.1,
		bar width=0.27cm, 
		legend style={at={(0.15,1)},anchor=north,legend columns=-1},
		ymajorgrids=true,
		ylabel={Accuracy (\%)},
		xlabel=Probability of mutation,
		xticklabels = {0.1,0.2,0.3,0.4,0.5,0.6,0.7,0.8,0.9},
		]
		\addplot[black,fill=lightBlue] table[x index=0,y index=1] \dataCLeft; 
		\addplot[black,fill=darkBlue] table[x index=0,y index=2] \dataCLeft; 
		\end{axis}
		\end{tikzpicture}
		\hspace{0.3cm}
		\begin{tikzpicture}
		font=\footnotesize
		\begin{axis}[
		ybar=0pt,
		width=8.0cm,height=5.5cm, 
		axis x line*=bottom,
		axis y line*=left,
		ytick align=outside,
		bar width=0.27cm, 
		enlarge x limits=0.1,
		ylabel near ticks,
		ymajorgrids=true,
		xtick=data,
		ylabel=Number of SNPs,
		xlabel=Probability of mutation,
		xticklabels = {0.1,0.2,0.3,0.4,0.5,0.6,0.7,0.8,0.9},
		]
		\addplot[black,fill=lightBlue] table[x index=0,y index=1] \dataCRight; 
		\addplot[black,fill=darkBlue] table[x index=0,y index=2] \dataCRight; 
		\end{axis}
		\end{tikzpicture}
	}
	\\
	\subfloat[\label{fig:05-d} IG+Proposed GA]{
		\begin{tikzpicture}
		font=\footnotesize 
		\begin{axis}[
		ybar=0pt,
		ylabel near ticks,
		width=8.0cm,height=5.5cm,
		axis x line*=bottom,
		axis y line*=left,
		ytick align=outside,
		xtick=data,
		enlarge x limits=0.1,
		bar width=0.27cm, 
		legend style={at={(0.15,1)},anchor=north,legend columns=-1},
		ymajorgrids=true,
		ylabel={Accuracy (\%)},
		xlabel=Probability of mutation,
		xticklabels = {0.1,0.2,0.3,0.4,0.5,0.6,0.7,0.8,0.9},
		]
		\addplot[black,fill=lightBlue] table[x index=0,y index=1] \dataDLeft; 
		\addplot[black,fill=darkBlue] table[x index=0,y index=2] \dataDLeft; 
		\end{axis}
		\end{tikzpicture}
		\hspace{0.3cm}
		\begin{tikzpicture}
		font=\footnotesize
		\begin{axis}[
		ybar=0pt,
		width=8.0cm,height=5.5cm, 
		axis x line*=bottom,
		axis y line*=left,
		ytick align=outside,
		bar width=0.27cm, 
		enlarge x limits=0.1,
		ylabel near ticks,
		ymajorgrids=true,
		xtick=data,
		ylabel=Number of SNPs,
		xlabel=Probability of mutation,
		xticklabels = {0.1,0.2,0.3,0.4,0.5,0.6,0.7,0.8,0.9},
		]
		\addplot[black,fill=lightBlue] table[x index=0,y index=1] \dataDRight; 
		\addplot[black,fill=darkBlue] table[x index=0,y index=2] \dataDRight; 
		\end{axis}
		\end{tikzpicture}
	}
	\caption{Classification accuracies and numbers of selected SNPs resulted from using a range of $P_m$ values.}
	\label{fig:05}
\end{figure}

%
\subsection{Classification accuracy and number of selected SNPs}
\label{method:exp1}
%
The resulting average classification accuracies and the number of selected SNPs are summarised in Table~\ref{table:2}. It can be seen that all of the methods used were competitive. The best method was IG+Proposed GA+FFS that provided a classification accuracy of 94.62~\% and 94.80~\% for linear and RBF kernels, respectively, showing that the proposed approach was able to achieve a better classification accuracy to that of using the entire features from the dataset (92.46~\%) while using far fewer features--only 0.86~\% of SNPs. The worst method was IG+GA+FFS that exhibited a classification accuracy of 76.62~\% for linear kernel and 77.08~\% for RBF kernel, markedly lower than any other methods. The reason for this might be that it provided too small a number of SNPs (21 SNPs) to be able to make effective classification. It can also be seen that the number of SNPs selected by wrapper methods, GA and Proposed GA, were still too high, 49.70~\% and 7.09~\% in linear case, respectively. When a filter method, IG, was used in combination with the wrapper methods, the number of selected SNPs reduced dramatically. For instance, IG+GA and IG+Proposed GA were able to reduce the number of features to 1.00~\% and 1.44~\% of the entire features in linear case while IG alone was able to reduce it to 1.98~\%. However, using too small number of SNPs could lead to a drop in performances as of IG+GA case. When FFS was added, the numbers of selected SNPs were further reduced: IG+FFS, IG+GA+FFS, and IG+Proposed GA+FFS were able to reduce the number of selected SNPs to 1.22~\%, 0.31~\%, and 0.86~\% of the entire SNPs in the dataset, respectively. Fortunately, the accuracies were improved in most cases except IG+GA+FSS case--using too small number of features (21 SNPs). The best performer was IG+Proposed GA+FFS which was able to select only 142 SNPs from the total of 16,579 SNPs.

\begin{table}[htbp]
	\centering
	\caption{Average classification accuracy and selected SNPs achieved by each method of the proposed approach (the best values are in bold). It is noted that the numbers of SNPs used in methods with FFS are constants.
	}
	\label{table:2}
	\begin{tabular}{|l|l|l|l|l|}
		\hline
		\multirow{2}{*}{\textbf{Methods}} & \multicolumn{2}{c|}{\textbf{Accuracy (\%)}} & \multicolumn{2}{c|}{\textbf{\#SNP}} \\ \cline{2-5} 
		& \multicolumn{1}{c|}{\textbf{Linear}} & \multicolumn{1}{c|}{\textbf{RBF}} & \multicolumn{1}{c|}{\textbf{Linear}} & \multicolumn{1}{c|}{\textbf{RBF}} \\ \hline
		Entire SNPs & 92.46 $\pm$ 1.98 & 92.46 $\pm$ 1.98 &\makecell{16,579.00 \\ (100 \%)} & \makecell{16,579.00\\ (100 \%)} \\ \hline
		GA & 92.92 $\pm$ 2.08 & 92.62 $\pm$ 2.03 &\makecell{8,239.80 $\pm$  73.84\\ (49.70 \%)} &\makecell{8,245.10 $\pm$  56.99 \\(49.73 \%)} \\ \hline
		Proposed GA & 91.69 $\pm$ 3.18 & 92.15 $\pm$ 2.85 &\makecell{1,176.50 $\pm$  573.61 \\(7.09 \%)} & \makecell{1,113.10 $\pm$  488.32\\ (6.71 \%)} \\ \hline
		IG & 92.15 $\pm$ 2.11 & 92.15 $\pm$ 2.11 &\makecell{329.00 $\pm$  87.73\\ (1.98 \%)} &\makecell{329.00 $\pm$  87.73\\(1.98 \%)} \\ \hline
		IG+GA & 90.31 $\pm$ 2.81 & 91.85 $\pm$ 2.18 & \makecell{165.60 $\pm$  45.47\\ (1.00 \%)} & \makecell{165.30 $\pm$  5.12 \\(1.00 \%)} \\ \hline
		IG+Proposed GA & 92.15 $\pm$ 2.76 & 91.85 $\pm$ 2.91 &\makecell{238.10 $\pm$  6.71\\ (1.44 \%)} &\makecell{241.30 $\pm$  9.52 \\(1.46 \%)} \\ \hline
		IG+FFS & 94.15 $\pm$ 2.27 & 94.00 $\pm$ 2.34 & \makecell{202.00 \\(1.22 \%)} &\makecell{202.00 \\(1.22 \%)} \\ \hline
		IG+GA+FFS & 76.62 $\pm$ 4.49 & 77.08 $\pm$ 4.61 & \textbf{\makecell{21.00 \\(0.13 \%)}} & \textbf{\makecell{21.00\\ (0.13 \%)}} \\ \hline
		IG+Proposed GA+FFS & \textbf{94.62 $\pm$ 2.21} & \textbf{94.80 $\pm$ 2.08} &\makecell{142.00\\ (0.86 \%)} & \makecell{142.00\\ (0.86 \%)} \\ \hline
	\end{tabular}
\end{table}

\begin{figure}[htbp]
	\captionsetup[subfloat]{aboveskip=-5em,belowskip=-1em,labelformat=empty}
	\centering
	\subfloat[]{
		\begin{tikzpicture}
		font=\footnotesize
		\begin{axis}[
		ybar=0pt,
		legend image code/.code={
			\draw [#1] (0cm,-0.1cm) rectangle (0.25cm,0.1cm);
		},
		width=8.0cm,height=5.5cm, 
		axis x line*=bottom,
		axis y line*=left,
		ytick align=outside,
		bar width=0.6cm, 
		legend style={at={(0.2,1)},anchor=north,legend columns=-1},
		enlarge x limits=0.2,
		ylabel near ticks,
		ymajorgrids=true,
		xtick=data,
		ylabel=Accuracy (\%),
		xlabel=Methods,
		xticklabels = {$\textrm{M}_{1-\textrm{Single}}$, $\textrm{M}_{1-\textrm{Combined}}$, $\textrm{M}_{2-\textrm{Single}}$, $\textrm{M}_{2-\textrm{Combined}}$},
		]
		\addplot[black,fill=lightBlue] table[x index=0,y index=1] \dataLeft; 
		\addplot[black,fill=darkBlue] table[x index=0,y index=2] \dataLeft; 
		\legend{Linear,RBF}
		\end{axis}
		\end{tikzpicture}
	}
	\subfloat[]{
		\begin{tikzpicture}
		font=\footnotesize
		\begin{axis}[
		ybar=0pt,
		width=8.0cm,height=5.5cm, 
		axis x line*=bottom,
		axis y line*=left,
		ytick align=outside,
		bar width=0.6cm, 
		enlarge x limits=0.2,
		ylabel near ticks,
		ymajorgrids=true,
		xtick=data,
		ylabel=Number of SNPs,
		xlabel=Methods,
		xticklabels = {$\textrm{M}_{1-\textrm{Single}}$, $\textrm{M}_{1-\textrm{Combined}}$, $\textrm{M}_{2-\textrm{Single}}$, $\textrm{M}_{2-\textrm{Combined}}$},
		]
		\addplot[black,fill=lightBlue] table[x index=0,y index=1] \dataRight; 
		\addplot[black,fill=darkBlue] table[x index=0,y index=2] \dataRight; 
		\end{axis}
		\end{tikzpicture}
	}
	\caption{Classification accuracy and number of SNPs after processed by FFS with single set of features and combined set of features (M$_1$ is IG+GA+FFS; M$_2$ is IG+Proposed GA+FFS).}
	\label{fig:09}
\end{figure}
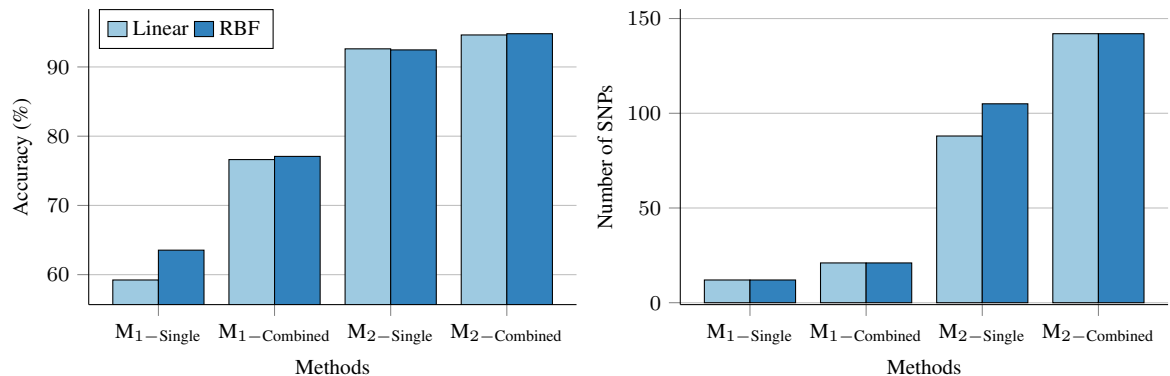

As mentioned earlier that, using features selected from either linear kernel or RBF kernel alone did not result in a good classification accuracy from IG+GA+FFS and IG+Proposed GA+FFS as show in Fig.~\ref{fig:09}. IG+GA+FFS and IG+Proposed GA+FFS could achieve at 59.23~\% and 92.62~\% accuracy with a set of features selected based on linear kernel, respectively, and at 62.54~\% and 92.46~\% accuracy based on RBF kernel, respectively. Thus, we used the unique features gleaned from the features selected by both kernels which resulted in a much better accuracy (76.62~\% and 94.62~\% for using IG+GA+FFS and IG+Proposed GA+FFS for using linear kernel, respectively, and 77.08~\%, 94.80~\% for using RBF kernel, respectively). It is clear that combining more relevant features led to a better performance. 
%
%
\pgfplotstableread{
	0	59.23076923 	63.53846154
	1 	76.62			77.08
	2	92.61538462		92.46153846
	3	94.62			94.80
}\dataLeft
\pgfplotstableread{
	0	12	12
	1	21	21
	2	88	105
	3	142 142
}\dataRight
%

%
\subsection{Results from an analysis of variance (ANOVA)}
\label{method:exp2}
%
One-Way ANOVA was used to test the hypotheses whether the average classification accuracies from different methods used were statistically different or not. In general, one-way ANOVA is used to compare more than two means whether at least a pair of the means are different or not. If it is so, a multiple comparison test will be used to find which pairs are significantly different.

As for the results of a one-way ANOVA analysis of the classification accuracies achieved by all of the methods, it was found that at least one pair of feature selection methods gave significantly different accuracies at $p\le0.05$, thus a multiple comparison was performed. The results from the multiple comparison show that the classification accuracies achieved by IG+GA+FFS with both linear and RBF kernels were significantly worse than those achieved by all of the other methods. In addition to this, the accuracy achieved by IG+Proposed GA+FFS with linear kernel was statistically significantly better than those achieved by IG+GA. As can be seen in Table~\ref{table:3}, even though these differences were slight (competitive to the others), IG+Proposed GA+FFS was able to reduce the number of selected features to 0.86~\%.
\begin{table}[htbp]
	\centering
	\footnotesize
	\setlength\tabcolsep{0.07cm}
	\caption{Results of pairwise comparison among all of the methods from the multiple comparison analysis (the significantly different accuracies are in bold).
	}
	\label{table:3}
	\begin{tabular}{|l|l|c|c|c|c|c|c|c|c|}
		\hline
		\multicolumn{1}{|c|}{\multirow{3}{*} { \textbf{Paired Method 1}}} & \multicolumn{1}{c|}{\multirow{3}{*}{\textbf{Paired Method 2}}} & \multicolumn{4}{c|}{\textbf{Linear kernel}} & \multicolumn{4}{c|}{\textbf{RBF kernel}} \\ \cline{3-10} 
		\multicolumn{1}{|c|}{} & \multicolumn{1}{c|}{} & \multirow{2}{*}{\textbf{\makecell{ Mean \\ difference }}} & \multirow{2}{*}{\textbf{$p$-value}} & \multicolumn{2}{c|}{\textbf{ \makecell{ 95\% confidence interval \\ for the mean difference} }} & \multirow{2}{*}{\textbf{\makecell{ Mean \\ difference}}} & \multirow{2}{*}{\textbf{$p$-value}} & \multicolumn{2}{c|}{\textbf{ \makecell{ 95\% confidence interval \\ for the mean difference} }} \\ \cline{5-6} \cline{9-10} 
		\multicolumn{1}{|c|}{} & \multicolumn{1}{c|}{} &  &  & \textbf{ \makecell{ Lower \\ Bound} } & \textbf{ \makecell{ Upper \\ Bound} } &  &  & \textbf{ \makecell{ Lower \\ Bound} } & \textbf{ \makecell{ Upper \\ Bound} } \\ \hline
		Entire SNPs & GA & -0.46 & 1.00 & -4.45 & 3.52 & -0.15 & 1.00 & -3.98 & 3.67 \\ \hline
		Entire SNPs & Proposed GA & 0.77 & 1.00 & -3.21 & 4.75 & 0.31 & 1.00 & -3.52 & 4.13 \\ \hline
		Entire SNPs & IG & 0.31 & 1.00 & -3.68 & 4.29 & 0.31 & 1.00 & -3.52 & 4.13 \\ \hline
		Entire SNPs & IG+GA & 2.15 & 0.73 & -1.83 & 6.14 & 0.62 & 1.00 & -3.21 & 4.44 \\ \hline
		Entire SNPs & IG+Proposed GA & 0.31 & 1.00 & -3.68 & 4.29 & 0.62 & 1.00 & -3.21 & 4.44 \\ \hline
		Entire SNPs & IG+FFS & -1.69 & 0.91 & -5.68 & 2.29 & -1.54 & 0.93 & -5.37 & 2.29 \\ \hline
		Entire SNPs & IG+GA+FFS & 15.85 & \textbf{0.00} & 11.86 & 19.83 & 15.38 & \textbf{0.00} & 11.56 & 19.21 \\ \hline
		Entire SNPs & IG+Proposed GA+FFS & -2.15 & 0.73 & -6.14 & 1.83 & -2.31 & 0.60 & -6.13 & 1.52 \\ \hline
		GA & Proposed GA & 1.23 & 0.99 & -2.75 & 5.21 & 0.46 & 1.00 & -3.37 & 4.29 \\ \hline
		GA & IG & 0.77 & 1.00 & -3.21 & 4.75 & 0.46 & 1.00 & -3.37 & 4.29 \\ \hline
		GA & IG+GA & 2.62 & 0.49 & -1.37 & 6.60 & 0.77 & 1.00 & -3.06 & 4.60 \\ \hline
		GA & IG+Proposed GA & 0.77 & 1.00 & -3.21 & 4.75 & 0.77 & 1.00 & -3.06 & 4.60 \\ \hline
		GA & IG+FFS & -1.23 & 0.99 & -5.21 & 2.75 & -1.38 & 0.96 & -5.21 & 2.44 \\ \hline
		GA & IG+GA+FFS & 16.31 & \textbf{0.00} & 12.32 & 20.29 & 15.54 & \textbf{0.00} & 11.71 & 19.37 \\ \hline
		GA & IG+Proposed GA+FFS & -1.69 & 0.91 & -5.68 & 2.29 & -2.15 & 0.69 & -5.98 & 1.67 \\ \hline
		Proposed GA & IG & -0.46 & 1.00 & -4.45 & 3.52 & 0.00 & 1.00 & -3.83 & 3.83 \\ \hline
		Proposed GA & IG+GA & 1.38 & 0.97 & -2.60 & 5.37 & 0.31 & 1.00 & -3.52 & 4.13 \\ \hline
		Proposed GA & IG+Proposed GA & -0.46 & 1.00 & -4.45 & 3.52 & 0.31 & 1.00 & -3.52 & 4.13 \\ \hline
		Proposed GA & IG+FFS & -2.46 & 0.57 & -6.45 & 1.52 & -1.85 & 0.83 & -5.67 & 1.98 \\ \hline
		Proposed GA & IG+GA+FFS & 15.08 & \textbf{0.00} & 11.09 & 19.06 & 15.08 & \textbf{0.00} & 11.25 & 18.90 \\ \hline
		Proposed GA & IG+Proposed GA+FFS & -2.92 & 0.33 & -6.91 & 1.06 & -2.62 & 0.43 & -6.44 & 1.21 \\ \hline
		IG & IG+GA & 1.85 & 0.86 & -2.14 & 5.83 & 0.31 & 1.00 & -3.52 & 4.13 \\ \hline
		IG & IG+Proposed GA & 0.00 & 1.00 & -3.98 & 3.98 & 0.31 & 1.00 & -3.52 & 4.13 \\ \hline
		IG & IG+FFS & -2.00 & 0.80 & -5.98 & 1.98 & -1.85 & 0.83 & -5.67 & 1.98 \\ \hline
		IG & IG+GA+FFS & 15.54 & \textbf{0.00} & 11.55 & 19.52 & 15.08 & \textbf{0.00} & 11.25 & 18.90 \\ \hline
		IG & IG+Proposed GA+FFS & -2.46 & 0.57 & -6.45 & 1.52 & -2.62 & 0.43 & -6.44 & 1.21 \\ \hline
		IG+GA & IG+Proposed GA & -1.85 & 0.86 & -5.83 & 2.14 & 0.00 & 1.00 & -3.83 & 3.83 \\ \hline
		IG+GA & IG+FFS & -3.85 & 0.07 & -7.83 & 0.14 & -2.15 & 0.69 & -5.98 & 1.67 \\ \hline
		IG+GA & IG+GA+FFS & 13.69 & \textbf{0.00} & 9.71 & 17.68 & 14.77 & \textbf{0.00} & 10.94 & 18.60 \\ \hline
		IG+GA & IG+Proposed GA+FFS & -4.31 & \textbf{0.02} & -8.29 & -0.32 & -2.92 & 0.28 & -6.75 & 0.90 \\ \hline
		IG+Proposed GA & IG+FFS & -2.00 & 0.80 & -5.98 & 1.98 & -2.15 & 0.69 & -5.98 & 1.67 \\ \hline
		IG+Proposed GA & IG+GA+FFS & 15.54 & \textbf{0.00} & 11.55 & 19.52 & 14.77 & \textbf{0.00} & 10.94 & 18.60 \\ \hline
		IG+Proposed GA & IG+Proposed GA+FFS & -2.46 & 0.57 & -6.45 & 1.52 & -2.92 & 0.28 & -6.75 & 0.90 \\ \hline
		IG+FFS & IG+GA+FFS & 17.54 & \textbf{0.00} & 13.55 & 21.52 & 16.92 & \textbf{0.00} & 13.10 & 20.75 \\ \hline
		IG+FFS & IG+Proposed GA+FFS & -0.46 & 1.00 & -4.45 & 3.52 & -0.77 & 1.00 & -4.60 & 3.06 \\ \hline
		IG+GA+FFS & IG+Proposed GA+FFS & -18.00 & \textbf{0.00} & -21.98 & -14.02 & -17.69 & \textbf{0.00} & -21.52 & -13.87 \\ \hline
	\end{tabular}
\end{table}
%
\subsection{Results from a PCA analysis}
\label{method:exp3}
%
After 142 most significant SNPs were selected, they were used to perform an analysis of the relationship between swine breeds by PCA. In general, principal component analysis (PCA) is a statistical procedure that uses an orthogonal transformation to convert a set of observations of possibly correlated variables into a set of values of linearly uncorrelated variables called principal components (PC). This transformation is defined in such a way that the first principal component (PC$_1$) has the largest possible variance. We performed PCA on both the entire SNP dataset and on the set of selected SNPs from our approach and compared the results. 

As is generally known, PC relates to the variance of data points in a dataset. PC$_1$ is the most significant PC and PC$_2$ is the second most significant PC. Calculated from the entire SNPs in the dataset, when PC$_1$ was plotted versus PC$_2$ as in Fig.~\ref{fig:10-a}, it can be seen that the data points representing each population of swine breed are closely grouped together and those representing different populations of swine breeds are clearly separated: Chinese pigs (Blue), landrace (Yellow), LargeWhite (Orange), Moura (Black), and Duroc (Red). These results are quite similar to those from a PCA analysis reported in~\cite{burgos-paz_porcine_2013}. That work used 206 village pig samples from the American continent including those from Canary Islands and Iberian Peninsula and 183 outgroup pigs from Iberian Peninsula, China, and some other global locations. Most samples were from Iberian Peninsula. The total number of samples was 389, and the total number of SNPs was 46,259. A conclusion was made that most of European village pigs were genetically similar; Chinese pigs--Jiangquhai, Jinhua, Meishan, and Xiang pigs--were distinctly dissimilar to breeds from other global locations; Landrace and LargeWhite were also genetically dissimilar to breeds from other global locations but more similar to Asian pigs than to wild boars and Iberian. Lastly, Duroc was also genetically distinctly different from other breeds. In this study, the dataset that we used had a smaller number of samples (356) as mentioned in Section~\ref{sec:dataset} and the number of selected SNPs used in the analysis was also much smaller (142), but the PCA analysis results achieved by the proposed approach as shown in Fig.~\ref{fig:10-b} are still almost the same as the PCA results achieved by using the entire SNPs in the dataset as well as the PCA results from~\cite{burgos-paz_porcine_2013}: namely, the data points for Landrace, Largewhite, and Moura were clearly separate from the data points for other breeds, and most distinctly separate from those of other breeds were data points for Duroc and Chinese pigs. These results demonstrate that the proposed approach is valid while providing a much higher computational efficiency than using the entire SNPs from the dataset.
%
%
\begin{figure}[htbp]
	\captionsetup[subfloat]{aboveskip=-5em,belowskip=-1em}
	\centering
	\ref{pcaLegend}
	\subfloat[\label{fig:10-a}Entire SNPs]{
	\begin{tikzpicture} 
	\begin{axis}[
	width=16cm,height=9cm,
	mark size = 2pt,
	axis x line*=bottom,
	axis y line*=left,
	ytick align=outside,
	xtick align=outside,
	grid=major,
	xlabel={PC$_1$},
	ylabel={PC$_2$},
	ylabel shift = -0.3cm,
	xlabel near ticks,
	legend style={
		anchor = south,
		column sep = 0pt,
		legend columns = 7,
		at = {(0.9\textwidth,1.1)},
	},
	legend to name=pcaLegend,
	legend cell align={left},
	]
	\addplot [
	scatter,
	only marks,
	point meta=explicit symbolic,
	scatter/classes={
		a={mark=*,xNevy},
		b={mark=*,xPurple1},
		c={mark=*,xPurple2},
		d={mark=*,xBlue1},
		e={mark=*,xBlue2},
		f={mark=*,xBlue3},
		g={mark=*,xBlue4},
		h={mark=*,xBlack},
		i={mark=*,xRed},
		j={mark=*,xYellow},
		k={mark=*,xOrange},
		l={mark=*,xGreen},
		m={mark=*,xGreen},
		n={mark=*,xTeal},
		o={mark=*,xMaroon},
		p={mark=*,xMint},
		q={mark=*,xOlive},
		r={mark=*,xCoral},
		s={mark=*,xMagenta},
		t={mark=*,xCyan},
		u={mark=*,xGray} 
	},
	] table [meta=label, row sep=\\] {
		x    y      label\\
	-20.38225755	-1.775850417	s	\\
	-19.18746304	0.687635191	s	\\
	-22.1375831	-9.234446535	s	\\
	-20.4625388	-7.747482027	s	\\
	-22.21821072	-8.842277127	s	\\
	-21.80037822	-9.700662593	s	\\
	-22.68730246	-8.738379867	s	\\
	-23.03305641	-7.871477757	s	\\
	-17.92104038	-4.867292313	s	\\
	-20.15137133	-3.729765155	s	\\
	-21.08503059	-8.613979515	s	\\
	-22.29417026	-6.908412352	s	\\
	-21.6198936	-3.420621977	s	\\
	-21.33094956	-7.314620211	s	\\
	-21.42346679	-6.088388817	s	\\
	-10.78809273	-14.19272281	r	\\
	-11.31381741	-12.37359821	r	\\
	-11.26166208	-15.60323016	r	\\
	-9.26295975	-14.07291995	r	\\
	-11.69516469	-14.07937887	r	\\
	-11.072276	-13.7849352	r	\\
	-11.25643398	-14.05582483	r	\\
	-12.12782133	-16.43425006	r	\\
	-10.08991697	-12.47797687	r	\\
	-11.47881517	-11.76830498	r	\\
	-11.53688668	-12.42407922	r	\\
	-11.41125966	-12.67488794	r	\\
	-12.55530993	-15.34719658	r	\\
	-12.41922112	-12.85792578	r	\\
	-7.654561867	-20.83867196	j	\\
	-7.548890657	-19.31319864	j	\\
	-8.600231875	-21.85914648	j	\\
	-8.232649752	-19.94656882	j	\\
	-8.479211282	-19.19527354	j	\\
	-8.123645496	-18.74929789	j	\\
	-8.314839228	-19.23824796	j	\\
	-6.910376714	-19.16482452	j	\\
	-9.214307726	-19.8502905	j	\\
	-10.88400895	-15.05852031	j	\\
	-8.450187802	-15.67488746	j	\\
	-7.394887196	-15.53866444	j	\\
	-9.730349748	-14.77673845	j	\\
	-10.34846727	-14.57259829	j	\\
	-9.723872256	-14.33518244	j	\\
	-8.185839087	-16.28130327	j	\\
	-11.17096274	-15.65922584	j	\\
	-6.208937486	-18.27523033	j	\\
	-7.680469528	-17.89105034	j	\\
	-8.198454964	-17.63959386	j	\\
	-5.975564136	-21.29677776	k	\\
	-6.37244504	-21.93126514	k	\\
	-5.452230046	-23.10956183	k	\\
	-5.156954838	-21.8648597	k	\\
	-6.455502158	-16.22158477	k	\\
	-5.949671403	-16.90335777	k	\\
	-5.199298161	-12.37194057	k	\\
	-4.638213502	-15.81643971	k	\\
	-8.320239882	-16.13375743	k	\\
	-7.269224033	-20.00693697	k	\\
	-8.70471608	-8.656577984	k	\\
	-7.563670585	-21.28912063	k	\\
	-7.237964738	-21.60635605	k	\\
	-6.560568732	-18.41679834	k	\\
	-6.162055475	-19.07362384	k	\\
	-7.148839625	-16.7981782	k	\\
	-7.455064078	-18.06332418	k	\\
	-6.542974894	-18.71770869	k	\\
	-7.373641187	-20.04911838	k	\\
	-8.569311888	-20.5643014	k	\\
	-12.73466541	50.74293395	i	\\
	-14.80864982	49.32882735	i	\\
	-13.53245903	51.87034304	i	\\
	-13.68246917	50.36401165	i	\\
	-12.94188631	47.99180449	i	\\
	-13.76332075	47.26400411	i	\\
	-14.78734271	46.46537748	i	\\
	-12.78958089	48.70588398	i	\\
	-14.53480753	45.39800433	i	\\
	-14.15599572	49.18822976	i	\\
	-12.02729761	46.69703152	i	\\
	-11.86430242	46.64987235	i	\\
	-13.54479009	48.38640594	i	\\
	-12.58601939	50.86043122	i	\\
	-13.72326945	49.11701451	i	\\
	-14.19820633	50.00729106	i	\\
	-13.21717894	47.46357655	i	\\
	-13.10141142	48.10745665	i	\\
	-12.91249607	48.14093091	i	\\
	-14.25725332	48.28114323	i	\\
	57.6229475	-0.679930464	f	\\
	56.52925302	-1.26962566	f	\\
	58.00163908	-0.638458822	f	\\
	57.62539581	-1.220565419	f	\\
	57.87695958	-0.851464144	f	\\
	58.04353521	-0.875761753	f	\\
	57.59932246	0.304948501	f	\\
	56.89126136	-0.770355354	f	\\
	59.02835288	1.012740673	f	\\
	56.26326196	-0.059293653	f	\\
	60.65890334	-0.042047884	f	\\
	59.14426361	-0.589654734	f	\\
	59.94875258	-0.463968173	f	\\
	60.87327033	-0.45887385	f	\\
	61.02952445	0.615199372	f	\\
	60.57994368	-0.469324994	f	\\
	53.26792479	-0.322427173	g	\\
	52.01111514	0.527538235	g	\\
	51.70029044	0.140779782	g	\\
	52.59431152	-0.527306515	g	\\
	53.27061437	-0.223337922	g	\\
	51.47286998	-0.440259309	g	\\
	53.45361891	-0.036340172	g	\\
	53.04061693	0.068009538	g	\\
	52.74857645	0.246559234	g	\\
	54.07301889	-0.250270062	g	\\
	54.47498213	0.81300741	g	\\
	64.61518699	3.994733957	e	\\
	63.93753215	4.11401859	e	\\
	60.83402972	4.136205454	e	\\
	63.97673238	4.623641353	e	\\
	63.6248729	4.087041986	e	\\
	63.99660152	4.376181653	e	\\
	64.01046628	3.7584905	e	\\
	63.96006834	3.58687317	e	\\
	64.65489126	3.759856735	e	\\
	64.19270252	4.331935591	e	\\
	63.7775955	3.049285043	e	\\
	62.90173588	4.27100874	e	\\
	64.29878598	3.748999406	e	\\
	65.0000811	3.936315272	e	\\
	64.55348457	4.041901307	e	\\
	63.92915738	3.724607908	e	\\
	56.11164891	-1.815820035	d	\\
	52.63458264	-2.579410823	d	\\
	54.74094681	-2.562695185	d	\\
	55.51967065	-1.590269071	d	\\
	55.83513343	-1.879107791	d	\\
	54.9510598	-2.270098752	d	\\
	53.90900536	-2.241886818	d	\\
	54.30814013	-2.648039823	d	\\
	53.46898656	-1.735244014	d	\\
	57.2729939	-1.678033226	d	\\
	56.16889911	-1.798301529	d	\\
	-12.63545413	-4.296305897	u	\\
	-13.46960384	-2.955097322	u	\\
	-13.53477173	-3.736297781	u	\\
	-13.4650268	-4.279876109	u	\\
	-12.76201564	-3.552253745	u	\\
	-13.34294906	-2.820006383	u	\\
	-13.83092125	-3.721158654	u	\\
	-12.95315975	-4.04762942	u	\\
	-12.43597418	-3.465488025	u	\\
	-11.50131802	-2.58627763	u	\\
	-14.05233655	-3.374236083	u	\\
	-13.79066581	-3.889615091	u	\\
	-12.81055327	-4.268521453	u	\\
	-13.34150714	-4.685489531	u	\\
	-15.25818451	-0.742535265	l	\\
	-14.3801865	-3.501165424	l	\\
	-15.70562268	-1.208337851	l	\\
	-12.31112001	2.445055239	l	\\
	-13.24960704	0.791914711	l	\\
	-12.3176818	-0.466438546	l	\\
	-12.21241662	4.192536096	l	\\
	-12.45997457	2.62715632	l	\\
	-13.19048238	2.194177716	l	\\
	-13.63744628	3.113774903	l	\\
	-8.204988571	1.485563352	a	\\
	-6.256557889	1.156081224	a	\\
	-7.425533598	0.07689795	a	\\
	-7.893229338	2.148962403	a	\\
	-6.738218912	-0.136166825	a	\\
	-9.049201671	1.90415624	a	\\
	-9.719110246	2.043632938	a	\\
	-7.729081639	1.19285371	a	\\
	-6.179068003	0.271489863	a	\\
	-7.839696894	2.224126817	a	\\
	-6.847825954	1.843523531	a	\\
	-6.776403554	1.536433232	a	\\
	-5.659034095	1.937443031	a	\\
	-10.01885465	1.110775336	a	\\
	-9.607871872	3.705626423	a	\\
	-8.959025661	5.108483635	a	\\
	-8.518697658	2.34020771	a	\\
	-10.54972042	3.914530681	a	\\
	-7.175509092	3.041639996	a	\\
	-8.364430326	4.106253354	a	\\
	-7.220631856	2.065082119	a	\\
	-1.892110915	-1.866738571	a	\\
	-1.386238446	0.863041273	a	\\
	-2.626523586	1.972519216	a	\\
	-4.287535452	-0.672270844	a	\\
	-3.812283497	0.619974723	a	\\
	-11.22662632	11.14511358	a	\\
	-11.65407961	7.24584355	a	\\
	-10.75696924	9.131014623	a	\\
	-8.923120349	8.262281118	a	\\
	-9.066983543	6.799910017	a	\\
	-11.41350865	9.97823811	a	\\
	-10.63992528	6.447161008	a	\\
	-9.755740607	8.48968632	a	\\
	-10.22843304	7.277472517	a	\\
	-11.70724388	12.63338659	a	\\
	-11.78079726	8.557927586	a	\\
	-11.14759895	11.18154512	a	\\
	-7.037256452	4.392079846	p	\\
	-6.823701953	2.706952554	p	\\
	-6.92162493	4.889180578	p	\\
	-7.36818361	3.926984159	p	\\
	-6.592852873	3.399721399	p	\\
	-7.411848034	4.362298738	p	\\
	-6.125133535	3.712650321	p	\\
	-11.33388692	-4.568249261	a	\\
	-10.48612339	-4.477092888	a	\\
	-11.16517746	-4.094336671	a	\\
	-10.66456111	-3.880450967	a	\\
	-10.05276241	-3.364339601	a	\\
	-10.27575236	-2.649592403	a	\\
	-11.80333253	0.900540187	a	\\
	-12.01228423	0.550422313	a	\\
	-12.13006381	0.568525217	a	\\
	-10.42304203	0.33334773	a	\\
	-12.64772508	0.226911627	a	\\
	-12.13639273	2.219342661	a	\\
	-13.66140273	1.822956685	a	\\
	-12.24423936	-0.28614339	a	\\
	-12.82474337	3.138849369	t	\\
	-12.07636752	3.562694819	t	\\
	-11.33749742	1.58576806	t	\\
	-12.11921227	1.924704903	t	\\
	-11.49729838	3.167397821	t	\\
	-12.07021962	1.198035505	t	\\
	-12.47962818	1.791287952	t	\\
	-13.30042567	3.183960434	t	\\
	-11.87749255	2.588793708	t	\\
	-11.57877474	2.785921174	t	\\
	-13.50487812	2.525757371	t	\\
	-12.73459141	3.89101802	t	\\
	-13.07325271	2.691906617	t	\\
	-12.42834343	2.594571661	t	\\
	-12.43718254	2.112593285	t	\\
	-16.48602288	3.362067945	o	\\
	-5.444656937	8.739433853	o	\\
	-7.647730902	3.569790695	o	\\
	-7.222616197	2.381881851	o	\\
	-6.075385299	4.056891078	o	\\
	-5.915008774	4.10236657	o	\\
	-5.983355428	5.577474724	o	\\
	-5.61024376	4.840023312	o	\\
	-5.492183437	4.509433299	o	\\
	-3.053128692	-7.64026482	r	\\
	-2.781232951	-7.910937479	r	\\
	-4.023125934	3.890429718	r	\\
	-1.81485266	-6.940202808	r	\\
	-0.164136472	-3.968989682	r	\\
	-3.28058579	-4.259871629	r	\\
	-1.374687884	-4.879627394	r	\\
	-1.337111085	-5.719527914	r	\\
	-0.66365997	-5.649092207	r	\\
	-1.651305917	-5.508521664	r	\\
	-6.494117159	19.79915738	h	\\
	-7.022652048	17.88643483	h	\\
	-8.521226324	21.93200486	h	\\
	-8.049690805	19.53985806	h	\\
	-7.312417472	16.65381888	h	\\
	-8.162904228	13.02227098	h	\\
	-6.516132854	20.83654654	h	\\
	-7.075275616	18.76862623	h	\\
	-7.31595034	16.47533734	h	\\
	-7.388388385	-2.324336585	q	\\
	-8.951032764	-1.449794789	q	\\
	-10.77608109	8.48236969	q	\\
	-6.693280531	-1.894291837	q	\\
	-8.947259259	-1.797817614	q	\\
	-8.475245933	-2.153392454	q	\\
	-6.792576081	-3.434411359	q	\\
	-12.43491029	-2.687366278	a	\\
	-13.89629827	-2.433472228	a	\\
	-14.92511718	-3.767926346	a	\\
	-15.23899325	-5.641500827	a	\\
	-13.81601529	-2.184627645	a	\\
	-12.60327128	-0.991161827	a	\\
	-14.2119373	-1.469631608	a	\\
	-13.00816096	-2.117318378	a	\\
	-14.33858514	-1.732705274	a	\\
	-13.24143772	-2.839902217	a	\\
	-13.04579447	-2.668733288	a	\\
	-13.77767211	1.657392522	a	\\
	-13.56841105	-1.706496536	a	\\
	-13.20272903	-0.209226798	a	\\
	-14.92429181	-2.79594903	a	\\
	-15.90825566	-4.076371116	a	\\
	-7.699716886	-2.139217379	b	\\
	-0.827860559	-3.742219251	b	\\
	-3.554775579	-4.757310247	b	\\
	-2.885625232	-4.187235376	b	\\
	-4.210622345	-5.413114218	b	\\
	-2.856557935	-3.717564463	b	\\
	-2.771328987	-4.779056978	b	\\
	-3.609902095	-4.64830139	b	\\
	-2.695511425	-4.171322987	b	\\
	-13.27652656	-6.774890879	m	\\
	-16.48056083	-8.169530704	m	\\
	-14.76457513	-8.333665464	m	\\
	-15.39866701	-7.637863639	m	\\
	-15.8690423	-8.454899922	m	\\
	-14.51799915	-7.510606468	m	\\
	-14.24440944	-10.57884808	m	\\
	-14.51338204	-11.92656978	m	\\
	-13.82914884	-12.09937889	m	\\
	-14.52349183	-10.67260347	m	\\
	-13.48816391	-10.46042045	m	\\
	-15.74249664	-10.59306965	m	\\
	-13.68185328	-10.36659959	m	\\
	-14.76483441	-9.75275944	n	\\
	-13.54786087	-9.929851046	n	\\
	-15.13825161	-10.17135129	n	\\
	-14.48111135	-8.233228465	n	\\
	-13.17040579	-10.45691775	n	\\
	-14.22678466	-8.546356929	n	\\
	-13.40307271	-9.583464772	n	\\
	-13.77381505	-9.066868421	n	\\
	-13.88962699	-9.856718826	n	\\
	-13.49773123	-10.05638303	n	\\
	-12.77185184	2.624786847	a	\\
	-11.1411006	5.5362548	a	\\
	-11.27522183	2.464438756	a	\\
	-9.52464323	0.635601715	a	\\
	-7.68789044	0.504496409	c	\\
	-8.978950117	-1.087987799	c	\\
	-8.80204875	-0.859286831	c	\\
	-9.20441988	0.684148007	c	\\
	-7.934800381	1.464514445	c	\\
	-9.000309964	-1.841547645	c	\\
	-9.920860194	-1.926104334	c	\\
	-9.500475581	0.032224541	c	\\
	-8.321479404	-2.124728214	c	\\
	-9.425121227	-2.152897128	c	\\
	-14.47822581	4.316579957	a	\\
	-13.37170717	2.119116853	a	\\
	-12.23508051	-1.141148331	a	\\
	-6.247756185	-2.135017536	a	\\
	-4.956490498	0.031355077	a	\\
	-4.560535088	-2.725679944	a	\\
	-4.97187454	-2.066126576	a	\\
	-9.568816875	-5.991371702	a	\\
	-4.305955137	-0.864149338	a	\\
	-4.631912934	-2.331636014	a	\\
	-4.746307123	0.321593164	a	\\
	-7.210194467	-1.314479087	a	\\
	-4.378425288	-0.423199275	a	\\
	-8.092030536	-1.112067704	a	\\
	-8.181849716	-1.836449739	a	\\
	-12.12847739	-0.326611865	a	\\
	-11.26045345	-1.857282008	a	\\
	-11.52569635	-1.654770694	a	\\
	};
	\legend{Creole,Piau,Zungo,Jiangquhai,Jinhua,Meishan,Xiang pig,Moura,Duroc,Landrace,Largewhite,Semi- feral,Wild boar,Yucatan,Hairless
		,Cuino,Ossabaw pig,Monteiro,Iberian,Hampshire,Guinea hog
	}
	\end{axis}
	\end{tikzpicture}
	}
	\\
	\subfloat[\label{fig:10-b}142 SNPs]{
	\begin{tikzpicture} 
	\begin{axis}[
	width=16cm,height=9cm,
	mark size = 2pt,
	axis x line*=bottom,
	axis y line*=left,
	ytick align=outside,
	xtick align=outside,
	grid=major,
	ylabel shift = -0.2cm,
	xlabel={PC$_1$},
	ylabel={PC$_2$},
	]
	\addplot [
	scatter,
	only marks,
	point meta=explicit symbolic,
	scatter/classes={
	a={mark=*,xNevy},
	b={mark=*,xPurple1},
	c={mark=*,xPurple2},
	d={mark=*,xBlue1},
	e={mark=*,xBlue2},
	f={mark=*,xBlue3},
	g={mark=*,xBlue4},
	h={mark=*,xBlack},
	i={mark=*,xRed},
	j={mark=*,xYellow},
	k={mark=*,xOrange},
	l={mark=*,xGreen},
	m={mark=*,xGreen},
	n={mark=*,xTeal},
	o={mark=*,xMaroon},
	p={mark=*,xMint},
	q={mark=*,xOlive},
	r={mark=*,xCoral},
	s={mark=*,xMagenta},
	t={mark=*,xCyan},
	u={mark=*,xGray} 
	},
	] table [meta=label,row sep=\\]{
		x    y      label\\
-3.487454066	0.091057052	s	\\
-3.733944995	0.24847142	s	\\
-4.786558997	0.178351693	s	\\
-3.646876064	1.667178744	s	\\
-4.793873394	0.214404184	s	\\
-4.123512063	1.227015345	s	\\
-5.250972611	2.23250276	s	\\
-5.074283949	0.451768812	s	\\
-4.671892306	-0.066695796	s	\\
-4.649937676	0.73957681	s	\\
-4.977353842	0.671326407	s	\\
-4.738999594	0.302317065	s	\\
-3.996932882	1.138128058	s	\\
-4.210907447	1.328898198	s	\\
-4.95430108	0.410324011	s	\\
-0.827597883	1.742521559	r	\\
-1.653943729	0.785403359	r	\\
-1.267953242	2.236528266	r	\\
-0.166692787	1.034586839	r	\\
-2.055124426	1.548783585	r	\\
-1.988097043	1.362170382	r	\\
-2.221423593	0.830451346	r	\\
-2.434999267	2.173137273	r	\\
-1.802401957	1.41313874	r	\\
-2.07878261	1.473103859	r	\\
-2.517314949	1.380818519	r	\\
-2.317801987	0.629114875	r	\\
-0.895277712	1.828109518	r	\\
-2.265459819	1.318379537	r	\\
-0.364730713	3.430910584	j	\\
-0.831959867	3.573000071	j	\\
-0.793323257	3.226952277	j	\\
-0.257406638	3.643767369	j	\\
-1.293431747	3.034682622	j	\\
-0.918030026	3.094435403	j	\\
-0.175012504	3.499701299	j	\\
0.050075706	2.833349577	j	\\
-0.647372706	1.850272185	j	\\
-1.547782812	2.651035828	j	\\
-1.449710779	2.972460312	j	\\
-0.992447519	2.260150322	j	\\
-1.422431793	2.638836012	j	\\
-2.128463368	2.555961415	j	\\
-2.58417868	2.332220618	j	\\
-1.243204762	2.854887899	j	\\
-1.615817792	1.832698373	j	\\
-0.684684998	4.208676415	j	\\
-1.522452958	3.182397008	j	\\
-1.225741811	3.522308363	j	\\
-1.257941392	1.162975508	k	\\
-0.61753328	1.112182589	k	\\
-1.289695823	2.156168406	k	\\
-1.177874846	0.763916701	k	\\
-0.095845741	1.64534422	k	\\
-0.013670811	2.027553875	k	\\
-1.038631104	2.027532879	k	\\
-0.877363042	2.251572376	k	\\
-1.857984468	1.722162173	k	\\
-1.231829304	1.607531671	k	\\
-0.222808191	0.58012091	k	\\
-0.417917215	2.316322394	k	\\
-0.794184133	2.905136589	k	\\
0.182011588	3.074991314	k	\\
-0.948912749	2.758328104	k	\\
-0.741824262	1.293823343	k	\\
-1.038400956	1.286601003	k	\\
-0.93701966	2.408674606	k	\\
-1.000568516	1.490646644	k	\\
-0.506054952	2.889344902	k	\\
-2.978682833	-6.194231317	i	\\
-2.918606527	-6.788674102	i	\\
-3.540787666	-7.010278257	i	\\
-2.31378856	-6.603607302	i	\\
-2.929364091	-6.77069249	i	\\
-3.273283972	-6.271322923	i	\\
-3.475377562	-6.239615482	i	\\
-2.757747328	-7.102803202	i	\\
-2.803446331	-5.450984246	i	\\
-2.052528467	-6.508551722	i	\\
-3.058881114	-6.968870813	i	\\
-2.673782649	-6.006297911	i	\\
-2.350394413	-6.474997158	i	\\
-1.722658755	-5.648266232	i	\\
-2.1923671	-6.826654778	i	\\
-1.723642635	-6.077179352	i	\\
-2.28340629	-6.664637971	i	\\
-2.503855517	-6.393505937	i	\\
-1.953599171	-6.467821306	i	\\
-2.439856024	-7.212362691	i	\\
11.12717448	0.226678227	f	\\
11.52568481	0.010578785	f	\\
11.10678321	0.332489083	f	\\
11.49954159	0.374924575	f	\\
11.48246607	0.001014943	f	\\
11.83880373	-0.337500311	f	\\
11.57349312	-0.558190679	f	\\
10.77196355	0.612192913	f	\\
12.12459341	0.07930997	f	\\
11.49248566	-0.392115428	f	\\
12.42571666	-0.581836731	f	\\
11.95401346	-0.552512517	f	\\
12.20230395	-0.373820353	f	\\
11.87818207	-0.313217808	f	\\
12.20876458	-0.19685316	f	\\
12.51531119	0.20683455	f	\\
10.1590793	-1.888523854	g	\\
9.858975729	-1.154224392	g	\\
10.6892645	-1.237996849	g	\\
10.68581582	-1.092754742	g	\\
10.36068404	-1.193849488	g	\\
10.67822851	-1.428651037	g	\\
10.40274075	-1.620885143	g	\\
10.65778023	-1.208550177	g	\\
10.346975	-1.382969606	g	\\
10.34556692	-1.507701023	g	\\
10.05992886	-2.118590336	g	\\
12.25789848	-0.528878047	e	\\
11.96391628	0.259125159	e	\\
12.03284978	0.108964727	e	\\
11.89974188	0.075577965	e	\\
12.0771835	-0.064560324	e	\\
12.20169058	-0.209002166	e	\\
12.3412896	-0.03728023	e	\\
11.87652557	0.002063738	e	\\
12.29896935	-0.108008584	e	\\
11.80340305	-0.156631449	e	\\
12.0503672	0.151053455	e	\\
12.34242952	0.300228587	e	\\
12.50538432	-0.193813897	e	\\
11.85394209	-0.276630744	e	\\
12.46696083	-0.124972386	e	\\
12.19958988	-0.23833546	e	\\
11.23087985	0.86964086	d	\\
9.947643677	1.072663117	d	\\
10.50409813	0.322682344	d	\\
11.09717333	0.574692632	d	\\
10.88576613	1.066046153	d	\\
10.95485622	0.920820264	d	\\
11.05841778	0.990787643	d	\\
10.10579333	0.519973753	d	\\
10.35901257	0.593355817	d	\\
11.18193419	0.281302391	d	\\
10.93287655	0.773646641	d	\\
-2.613541024	2.42182329	u	\\
-2.179324726	1.708954576	u	\\
-2.821163462	1.86811912	u	\\
-2.263944989	1.179721994	u	\\
-1.358213408	1.696728012	u	\\
-2.830020355	1.443816836	u	\\
-2.341753987	1.161150698	u	\\
-2.953622199	1.062723472	u	\\
-2.615291989	1.597383981	u	\\
-2.161204009	1.740913899	u	\\
-2.768463187	1.68921662	u	\\
-2.606338875	1.344103847	u	\\
-2.998370605	0.988242566	u	\\
-2.565737185	2.394940835	u	\\
-3.02305731	-0.907636448	l	\\
-3.269188059	0.646694667	l	\\
-2.793534121	-1.413722751	l	\\
-3.33611721	-0.658570848	l	\\
-1.485263049	-0.215079248	l	\\
-2.400760001	0.084607549	l	\\
-1.957756173	-0.582286986	l	\\
-2.276153104	-0.45460854	l	\\
-3.031252125	-0.506504317	l	\\
-2.223909384	-0.940160407	l	\\
-0.626820872	-0.521921435	a	\\
-1.16595498	-0.839950289	a	\\
-1.185097487	-1.045233794	a	\\
-1.679667808	-1.398824056	a	\\
-2.461018512	0.12512458	a	\\
-1.27840037	-2.181361041	a	\\
-2.473240923	-0.056507175	a	\\
-1.132118218	-0.361767893	a	\\
-1.957634366	-1.154238685	a	\\
-1.293219593	-0.116618616	a	\\
-1.394805664	-0.674543377	a	\\
-0.315477616	-0.614044727	a	\\
-0.488231074	-1.016056589	a	\\
-3.47283322	-0.655510915	a	\\
-2.312980665	-1.555331391	a	\\
-2.477682678	-1.002156812	a	\\
-1.894982145	0.754235724	a	\\
-2.58713691	-0.233822687	a	\\
-1.785319364	-1.47635642	a	\\
-2.354403009	-1.269924574	a	\\
-0.866289948	-0.111041624	a	\\
-0.493873996	-0.244460036	a	\\
-0.622964346	-0.09814351	a	\\
-1.279846143	0.960797626	a	\\
-0.827036194	-1.186971637	a	\\
-1.495156155	-0.035901745	a	\\
-3.08776043	-2.515347817	a	\\
-1.205103097	-1.476441432	a	\\
-2.483191015	-1.244135566	a	\\
-2.199375905	-1.086573666	a	\\
-2.394464439	-0.901103706	a	\\
-3.186199316	-1.435520687	a	\\
-2.445178838	-1.148326564	a	\\
-2.266887241	-0.883334456	a	\\
-3.225907123	0.017585815	a	\\
-3.196802904	-2.007193323	a	\\
-2.727252643	-0.089309332	a	\\
-3.550668693	-1.467120615	a	\\
0.108277562	0.609881883	p	\\
0.005610261	0.639531245	p	\\
-1.14305684	1.206899333	p	\\
-1.22366842	1.363511936	p	\\
-0.453369841	2.679112198	p	\\
-0.079510081	1.491653704	p	\\
-0.46811702	0.697729839	p	\\
-3.252121684	1.479784647	a	\\
-2.824614042	0.321967967	a	\\
-2.366551125	0.436957684	a	\\
-3.326482293	0.432547432	a	\\
-3.492815797	0.031730331	a	\\
-2.296067094	0.390254705	a	\\
-2.373647284	0.569190903	a	\\
-1.045260062	0.385003631	a	\\
-2.322802611	1.281710452	a	\\
-1.676761287	-0.169189924	a	\\
-1.498594121	0.965059314	a	\\
-2.746810651	-0.129366119	a	\\
-3.609610986	-0.426615942	a	\\
-1.727667438	0.333331855	a	\\
-2.454968967	1.238175686	t	\\
-2.762661566	1.77460338	t	\\
-1.685790144	0.436075296	t	\\
-3.715342138	0.753926222	t	\\
-1.772820359	2.012846618	t	\\
-4.145470994	0.924912425	t	\\
-3.500140717	0.947968944	t	\\
-2.33731524	1.479133815	t	\\
-3.074525608	1.18641365	t	\\
-2.64611773	2.034865538	t	\\
-2.903867349	1.364546079	t	\\
-2.70714251	1.553947337	t	\\
-2.322316453	1.823219962	t	\\
-2.485089468	0.864797952	t	\\
-2.716004398	0.959644768	t	\\
-2.632478952	-1.631525855	o	\\
-1.408081227	-0.716525891	o	\\
-1.275859855	0.224303516	o	\\
-0.308269712	0.900675503	o	\\
-0.054081381	0.92063459	o	\\
-0.147796806	-0.195035863	o	\\
-1.762451161	-0.67045588	o	\\
-0.167933265	-0.302813811	o	\\
-0.576574382	-0.278910439	o	\\
-0.661156145	0.266993945	r	\\
-0.008352996	0.886837529	r	\\
-1.365164452	0.722243036	r	\\
0.969338785	0.358060403	r	\\
-0.741442147	-0.518045592	r	\\
-1.803865936	-0.017338963	r	\\
-0.31863167	0.551914013	r	\\
-0.046949393	1.108721541	r	\\
-1.084477739	0.012595091	r	\\
-0.759442718	0.547491033	r	\\
-2.521017056	-3.258014991	h	\\
-1.425809174	-2.398379924	h	\\
-2.033996726	-3.303587819	h	\\
-1.202038481	-2.692615863	h	\\
-1.375256193	-2.504811995	h	\\
-2.298122637	-2.27695462	h	\\
-1.186296463	-2.773273531	h	\\
-2.44079296	-1.970323951	h	\\
-0.819888204	-2.463201487	h	\\
-0.792904367	-0.216459494	q	\\
1.160089192	-1.590992822	q	\\
-1.610588422	-1.573045297	q	\\
0.623660904	-0.176786495	q	\\
0.870239497	-0.460737474	q	\\
1.066645786	-0.420394193	q	\\
0.890478385	-0.051772566	q	\\
-3.353029317	0.126110716	a	\\
-4.123236253	0.835425992	a	\\
-4.596513322	1.284312365	a	\\
-2.316572136	-0.055762543	a	\\
-3.417199293	1.089116153	a	\\
-3.394034281	-0.763197415	a	\\
-3.435772378	0.872632869	a	\\
-1.749583449	0.37390874	a	\\
-3.798503414	-0.530910885	a	\\
-3.486398535	0.008504306	a	\\
-3.331177722	0.681311071	a	\\
-1.749579454	-1.251768062	a	\\
-2.841375947	1.109885002	a	\\
-3.539001229	-0.101331876	a	\\
-3.852011081	1.056855074	a	\\
-2.871514254	0.557675793	a	\\
-3.225310853	-0.591699368	b	\\
-0.623568926	-1.643853452	b	\\
-1.453733219	-1.623095382	b	\\
-1.216668196	-1.792218264	b	\\
-1.560242029	-1.36402776	b	\\
-0.9140663	-1.319489526	b	\\
-0.968718543	-0.733767156	b	\\
-0.893546752	-0.443544763	b	\\
-0.727585492	-1.779264441	b	\\
-2.251306839	2.39079862	m	\\
-1.840348836	2.349146155	m	\\
-2.182293184	2.84260078	m	\\
-2.327422156	1.451910123	m	\\
-3.65164436	0.994387796	m	\\
-2.356119727	2.486379878	m	\\
-1.555826449	2.571539396	m	\\
-2.001757736	2.815961271	m	\\
-0.881888703	3.324634289	m	\\
-1.614912108	2.860796099	m	\\
-1.574902537	2.41567136	m	\\
-1.981250548	3.281569771	m	\\
-2.811723207	2.099087063	m	\\
-2.572138733	0.111972878	n	\\
-1.855175838	0.562722813	n	\\
-2.43764118	0.264385496	n	\\
-2.623638166	0.777791651	n	\\
-2.741816321	0.614026442	n	\\
-3.396544012	0.216552691	n	\\
-1.911167228	1.840353582	n	\\
-3.670648951	0.382953267	n	\\
-2.337733908	0.515897448	n	\\
-1.760215682	0.173482569	n	\\
-3.05352251	-0.336326786	a	\\
-1.96584026	-1.969121393	a	\\
-2.85007159	-0.352410047	a	\\
-2.076554887	0.119640237	a	\\
-2.632332433	-1.142221184	c	\\
-2.936612548	-1.087653465	c	\\
-3.103610769	-0.88593796	c	\\
-2.851233661	-1.465547147	c	\\
-2.783555191	-0.44822168	c	\\
-3.043411441	-0.806633527	c	\\
-2.942068368	-0.669160284	c	\\
-2.354697037	0.564097003	c	\\
-3.186696784	-0.036932897	c	\\
-3.045233777	-0.01754536	c	\\
-3.634658952	-0.42390743	a	\\
-2.611459568	-1.324331648	a	\\
-4.046765012	0.491678864	a	\\
0.096993576	-0.282298779	a	\\
-1.983691567	-1.253583964	a	\\
-1.524795291	-1.660941463	a	\\
-0.697970837	-0.418284231	a	\\
-2.164016833	0.400874948	a	\\
-1.040040118	-0.926442797	a	\\
-0.447629841	-0.631000181	a	\\
-0.886764252	0.2198427	a	\\
-1.832131251	0.400630011	a	\\
-2.570356294	0.239927828	a	\\
-1.062958021	0.265320839	a	\\
-2.934267781	-0.027219033	a	\\
-4.843432554	0.780588705	a	\\
-3.410432142	-0.856355704	a	\\
-4.079704272	-0.131345667	a	\\
	};
	\end{axis}
	\end{tikzpicture}
	}
	\caption{Conventional PCA projection of SNPs in the dataset.}
	\label{fig:10}
\end{figure}
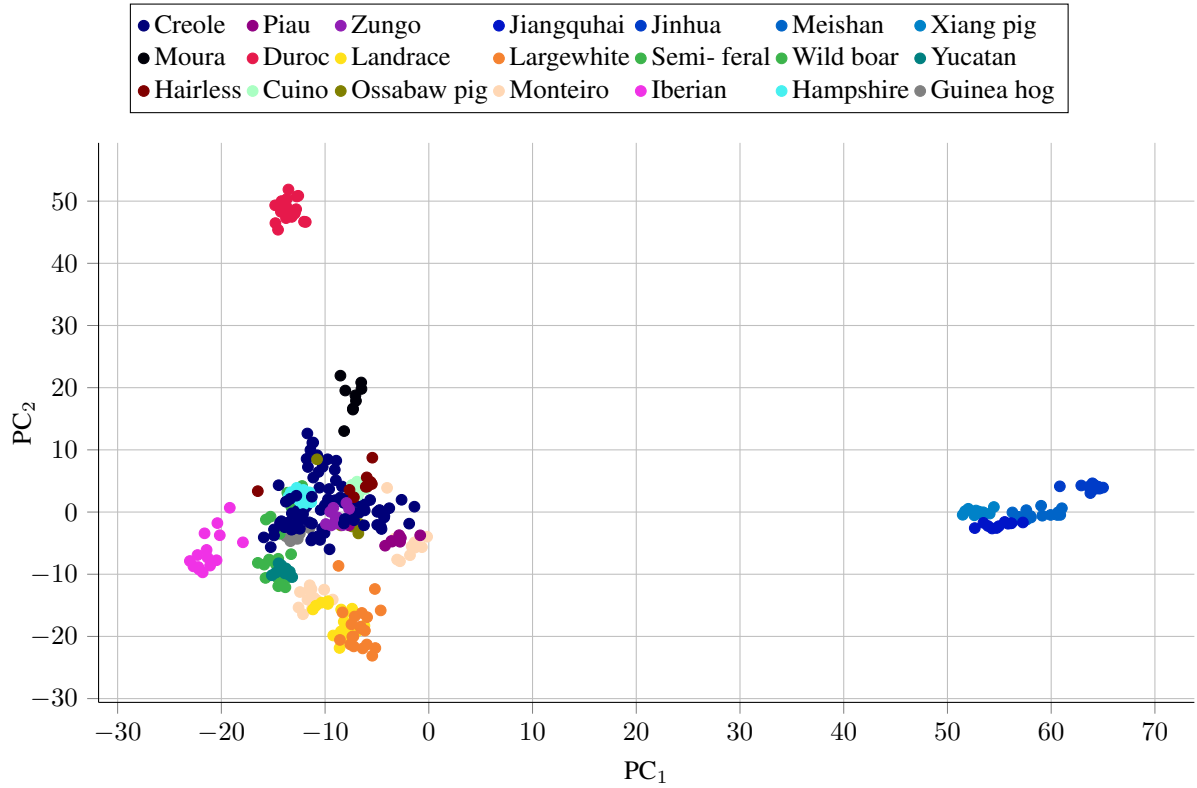
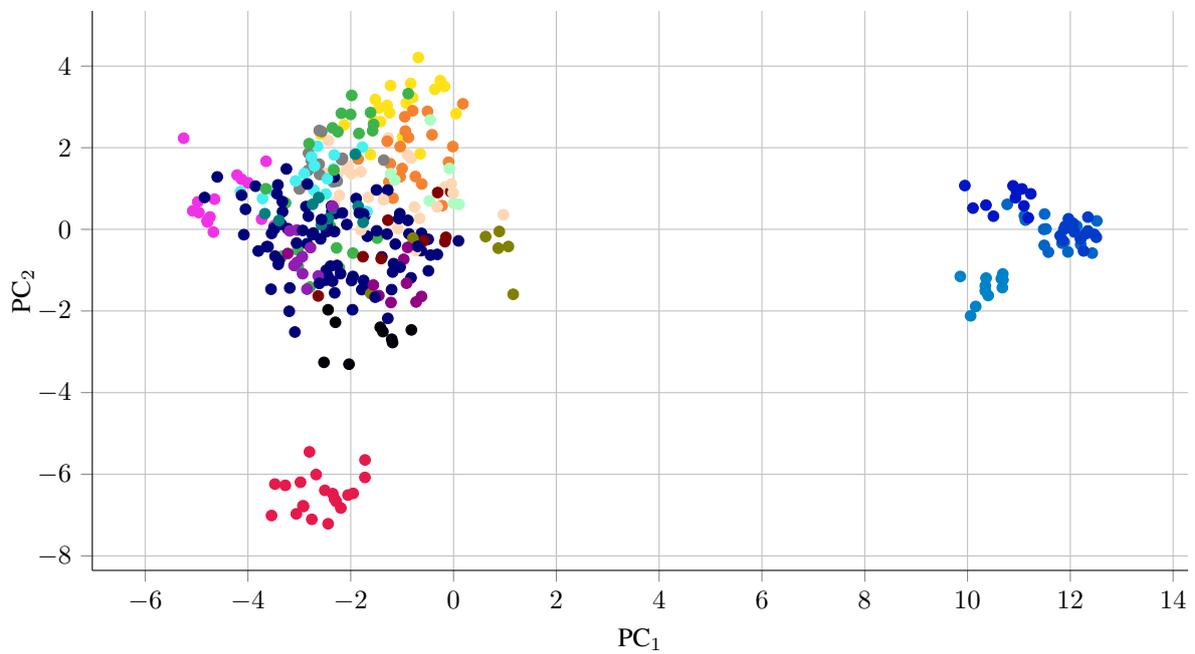
\begin{figure}[htbp]
	\centering
	\footnotesize
	\ref{lineLegend}
	\begin{tikzpicture}
	\begin{axis}[
	width=15.0cm,height=7cm, 
	xlabel=Generation,
	ylabel=Accuracy,
	enlarge x limits=0,
	ylabel near ticks,
	axis x line*=bottom,
	ytick align=outside,
	xtick align=outside,
	ymajorgrids=true,
	legend style={
		anchor = south,
		column sep = 0pt,
		legend columns = 4,
		at = {(0.95,0.5)},
	},
	legend to name=lineLegend,
	legend cell align={left},
	]
	\addplot[smooth,color=red,mark=*,red]
	plot coordinates {
		(0,89.29)
		(1,89.68)
		(2,89.68)
		(3,89.68)
		(5,89.68)
		(5,89.68)
		(6,89.68)
		(7,89.68)
		(8,89.68)
		(9,89.68)
		(10,89.68)
	};
\addlegendentry{GA(Linear)}
	\addplot[smooth,color=red,mark=square*,red]
	plot coordinates {
		(0,89.29)
		(1,89.68)
		(2,89.68)
		(3,89.68)
		(4,89.68)
		(5,89.68)
		(6,89.68)
		(7,89.68)
		(8,89.68)
		(9,89.68)
		(10,89.68)
	};
\addlegendentry{GA(RBF)}
	\addplot[smooth,color=green,mark=*,green]
	plot coordinates {
		(0,89.29)
		(1,89.64)
		(2,89.64)
		(3,89.67)
		(4,90.70)
		(5,90.70)
		(6,90.70)
		(7,90.70)
		(8,90.70)
		(9,90.70)
		(10,90.70)
	};
\addlegendentry{Proposed GA(Linear)}
	\addplot[smooth,color=green,mark=square*,green]
	plot coordinates {
		(0,89.29)
		(1,89.33)
		(2,89.95)
		(3,89.95)
		(4,89.95)
		(5,89.95)
		(6,90.64)
		(7,90.64)
		(8,90.64)
		(9,90.64)
		(10,90.64)
	};
\addlegendentry{Proposed GA(RBF)}
	\addplot[smooth,color=orange,mark=*,orange]
	plot coordinates {
		(0,93.07)
		(1,94.10)
		(2,94.10)
		(3,94.13)
		(4,94.13)
		(5,94.45)
		(6,94.45)
		(7,94.45)
		(8,94.45)
		(9,94.49)
		(10,94.49)
	};
\addlegendentry{IG+GA(Linear)}
	\addplot[smooth,color=orange,mark=square*,orange]
	plot coordinates {
		(0,93.07)
		(1,94.48)
		(2,94.48)
		(3,94.48)
		(4,94.48)
		(5,94.48)
		(6,94.48)
		(7,94.48)
		(8,94.48)
		(9,94.48)
		(10,94.48)
	};
\addlegendentry{IG+GA(RBF)}
	\addplot[smooth,color=blue,mark=*,blue]
	plot coordinates {
		(0,93.07)
		(1,93.43)
		(2,93.47)
		(3,93.47)
		(4,93.47)
		(5,93.81)
		(6,93.81)
		(7,93.81)
		(8,93.81)
		(9,93.81)
		(10,93.81)
	};
\addlegendentry{IG+Proposed GA(Linear)}
	\addplot[smooth,color=blue,mark=square*,blue]
	plot coordinates {
		(0,93.07)
		(1,94.44)
		(2,94.44)
		(3,94.44)
		(4,94.44)
		(5,94.45)
		(6,94.45)
		(7,94.45)
		(8,94.45)
		(9,94.45)
		(10,94.45)
	};
\addlegendentry{IG+Proposed GA(RBF)}
	\end{axis}
	\end{tikzpicture}
	\caption{The classification accuracy from each generation of the first randomly-seeded dataset.}
	\label{fig:08} 
\end{figure}
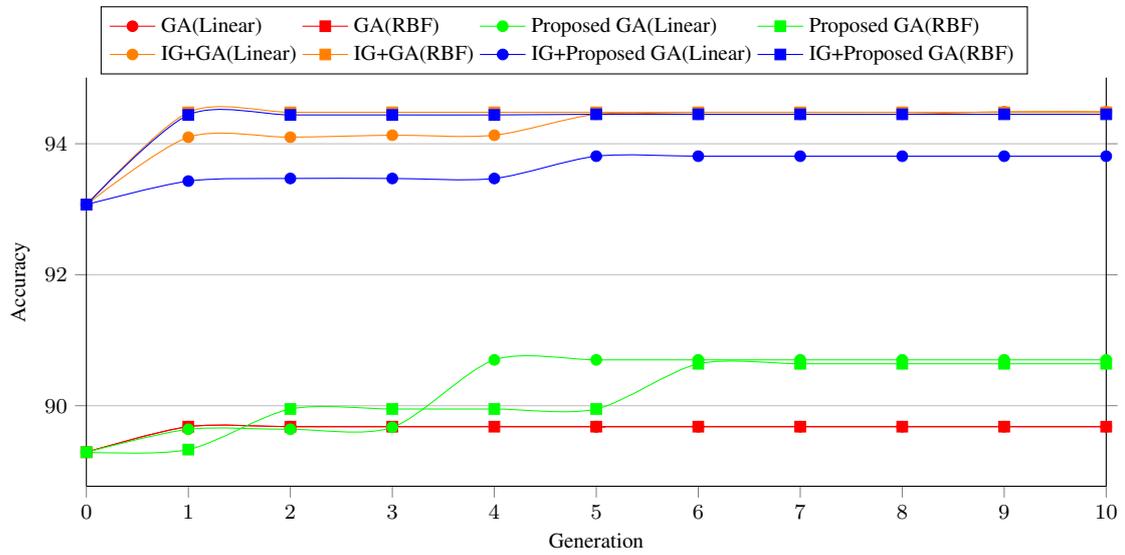
\pgfplotstableread{
	0	4.20	3
	1	6.80	6.60
	2 	6.40	3.50
	3	3.57	6.00
	
}\datayy
%
\section{CONCLUSION}
\label{sec:conclude}
%
A small number of usable Porcine SNPs for swine classification can be suitably selected by using feature selection and classification techniques. This study employed the methods of IG, GA, Proposed GA, IG+GA, IG+Proposed GA, IG+FFS, IG+GA+FFS, and IG+Proposed GA+FFS for finding a small number of suitable SNPs and SVM for classification. It was found that IG+Proposed GA+FFS was able to reduce the number of suitable SNPs to 0.86~\% of the total number of SNPs in the dataset used while provided a high classification accuracy of 94.80~\% that was higher than those achieved by the other methods. Compared to a classification result reported in the literature, the result from the proposed approach matched well with it, demonstrating the validity of the approach that also provides a much higher computational efficiency. A future work should be determining the genes that are related to these selected SNPs, finding their biological pathway, and determining the gene ontology annotation that relates to the genes. The information gained from this future work will be very useful in the biology field.
%
%
%
\bibliographystyle{IEEEtran}
\bibliography{References}

\end{document}